\documentclass[preprintnumbers,amsmath,11pt,amssymb,floatfix,superscriptaddress,
showpagenumber,showpacs,nofootinbib]{revtex4}
\usepackage{graphicx}
\usepackage{epsfig}
\usepackage{bm}
\usepackage{amsfonts}

\input epsf

\begin{document}

\title{\Large Holographic polytropic $f(T)$-gravity models}

\author{Surajit Chattopadhyay}
\email{surajitchatto@outlook.com, surajcha@iucaa.ernet.in}
\affiliation{ Pailan College of Management and Technology, Bengal
Pailan Park, Kolkata-700 104, India.}

\author{Abdul Jawad}
\email{abduljawad@ciitlahore.edu.pk, jawadab181@yahoo.com}
\affiliation{Department of Mathematics, COMSATS Institute of\\
Information Technology, Lahore-54000, Pakistan.}

\author{Shamaila Rani}
\email{drshamailarani@ciitlahore.edu.pk,
shamailatoor.math@yahoo.com}
\affiliation{Department of Mathematics, COMSATS Institute of\\
Information Technology, Lahore-54000, Pakistan.}

\date{\today}

\begin{abstract}
The present paper reports a study on the cosmological consequences
arising from reconstructing $f(T)$ gravity through new
holographic-polytropic dark energy. We assume two approaches, namely
a particular form of Hubble parameter $H$ and a solution for $f(T)$.
We obtain the deceleration parameter, effective equation of state as
well as torsion equation of state parameters from total density and
pressure in both cases. It is interesting to mention here that the
deceleration and torsion equation of state represent transition from
deceleration to acceleration phase. We study the statefinder
parameters under both approaches which result that statefinder
trajectories are found to attain $\Lambda$CDM point. The comparison
with observational data represents consistent results. Also, we
discuss the stability of reconstructed models through squared speed
of sound which represents stability in late times.
\end{abstract}


\maketitle

\section{Introduction}

The accelerated expansion of the universe is strongly manifested
after the discovery of unexpected reduction in the detected energy
fluxes coming from SNe Ia \cite{S27,S271}. Other observational data
like CMBR, LSS and galaxy redshift surveys \cite{S28,S281,S282} also
provide evidences in this favor. These observations propose a
mysterious form of force, referred as dark energy (DE), reviewed in
\cite{copeland-2006,bambareview,DErev,nojirireview}, which takes
part in the expansion phenomenon and dominates overall energy
density of the universe. This has two remarkable features: its
pressure must be negative in order to cause the cosmic acceleration
and it does not cluster at large scales. In spite of solid favor
about the presence of DE from the observations, its unknown nature
is the biggest puzzle in astronomy. In the last nineties, this
expansion was detected but the evidence for DE has been developed
during the past decade.

Physical origin of DE is one of the largest mysteries not only in
cosmology but also in fundamental physics \cite{copeland-2006,star1,satro1,star2,star3}.
The dynamical nature of DE can be originated from different models
such as cosmological constant, scalar field models, holographic DE
(HDE), Chayplygin gas, polytropic gas and modified gravity theories.
Various DE models are discussed in the references \cite{DE10,DE11,DE12,DE13,DE14,DE15,DE16,DE17}.
The modified theories of gravity are the generalized models came
into being by modifying gravitational part in general relativity
(GR) action while matter part remains unchanged. At large distances,
these modified theories, modify the dynamics of the universe. The
$f(R)$ theory is the modification of GR which modifies the Ricci
(curvature) scalar $R$ to a general differentiable function. The
gravitational interaction is established through curvature with the
help of Levi-Civita connection. There is another theory which is the
result of unification of gravitation and electromagnetism. It is
based on mathematical structure of absolute or distant parallelism,
also referred as teleparallelism which led to teleparallel gravity.
In this gravity, torsion is used as the gravitational field via
Witzenb\"{o}ck connection. The modification of teleparallel gravity
in the similar fashion of $f(R)$ gravity gives generalized
teleparallel gravity, $f(T)$ gravity where $f$ is general
differentiable function of torsion scalar.

The search for a viable DE model (representing accelerated expansion
of the universe) is the basic key leading to the reconstruction
phenomenon, particularly in modified theories of gravity
\cite{NO,NO1,NO2,NO3}. This reconstruction scheme works on the idea
of comparison of corresponding energy densities to obtain the
modified function in the underlying gravity. Daouda et al. \cite{S9}
developed the reconstruction scheme via HDE model in $f(T)$ gravity
and found that the reconstructed model may cross the phantom divide
line in future era. Setare and Darabi \cite{S10} assumed the scale
factors in power-law form and obtained well defined solutions.
Farooq et al. \cite{S11} reconstructed $f(T)$ model by taking
$(m,n)-$type HDE model and discussed its viability as well as
cosmography. They showed that this model is viable, compatible with
solar system test, ghost-free and has positive gravitational
constant. Karami and Abdolmaleki \cite{S13} obtained equation of
state (EoS) parameter for the reconstructed $f(T)$ models by taking
HDE, new agegraphic DE as well as their entropy-corrected versions
and found transition from non-phantom to phantom phase only in
entropy-corrected versions showing compatibility with the recent
observations. Sharif and Rani \cite{SJ,SJ3} explored this theory via
some scalar fields, nonlinear electrodynamics and entropy-corrected
HDE models and analyze the accelerated expansion of the universe.

Holographic DE models are widely used for explaining the present day
DE scenario and evolution of the universe. These are based on the
holographic principle which naively asks that the combination of
quantum mechanics and quantum gravity requires three-dimensional
world to be an image of data that can be stored on a two-dimensional
projection much like a holographic image \cite{S7,S71}. It is useful
to reveal the entropy bounds of black holes (BHs) which lead to the
formulation of the holographic principle. It is well established
that the area of a BH event horizon never decreases with time so
called area theorem. If a matter undergoes gravitational collapse
and converts into a BH, the entropy associated with the original
system seems to disappear since the final state is unique. This
process clearly violates the second law of thermodynamics. In order
to avoid this problem, Bekenstein \cite{T8z} proposed generalized
second law of thermodynamics on the basis of area theorem which is
stated as, BH carries an entropy proportional to its horizon area
and that the total entropy of ordinary matter system and BH never
decreases. Mathematically, it can be written as
\begin{equation}\label{1.5.1z}
\frac{dS_{tot}}{dt}\geq0.
\end{equation}
Here $S_{tot}=S+S_{BH},~S$ represents the entropy of matter (body)
outside a BH and $S_{BH}$ is the entropy of BH.

In the construction of HDE model, the relation between ultra-violet
(UV) ($\Lambda$) and infra-red (IR) ($L$) cutoffs as proposed by
Cohen et al. \cite{S8} plays a key role. It is suggested that for an
effective field theory in a box of size $L$ with $\Lambda$, the
entropy $S$ scales extensively, i.e., $S\sim L^3\Lambda^3$. However,
the maximum entropy in a box possessing volume $L^3$ (growing with
the area of the box) behaves non-extensively in the framework of BH
thermodynamics so called Bekenstein entropy bound. For any $\Lambda$
(containing sufficiently large volume), the entropy of effective
field theory will exceed the Bekenstein limit which can be satisfied
if we limit the volume of the system as follows
\begin{equation}\label{1.6.1z}
S=L^3\Lambda^3\lesssim S_{BH}\equiv\pi L^2M_p^2,
\end{equation}
where $S_{BH}$ has radius $L$.

It can be seen from the above inequality that IR cutoff $L$ (scales
as $\Lambda^{-3}$) is directly associated with UV cutoff and cannot
be chosen independently from it. Moreover, there occur some problems
in saturating the above inequality because Schwarzschild radius is
much larger than the box size and hence produces incompatibility
problem with effective field theory. To avoid this problem, Cohen et
al. \cite{S8} proposed a strong constraint on the IR cutoff which
excludes all states that lie within the Schwarzschild radius, i.e.,
\begin{equation}\label{1.6.2z}
L^3\Lambda^4=L^3\rho_{\Lambda}\lesssim LM^2_p.
\end{equation}
Here, left and right hand sides correspond to the total energy of
the system (since the maximum energy density in the effective field
theory is $\Lambda^4$) and mass of the Schwarzschild BH,
respectively. Also, IR cutoff $L$ is being scaled as $\Lambda^{-2}$
which is more restrictive limit than (\ref{1.6.1z}). The above
relation indicates that the maximum entropy of the system will be
$S^{\frac{3}{4}}_{BH}$. Li \cite{T11z} developed the energy density
for DE model by saturating the above inequality as follows
\begin{equation}\label{0.2.1z}
\rho_{\Lambda}=3\zeta^{2}M^{2}_pL^{-2},
\end{equation}
where $\zeta$ is the dimensionless HDE constant parameter. The
interesting feature of this density is that it provides a relation
between UV (bound of vacuum energy density) and IR (size of the
universe) cutoffs. However, a controversy about the selection of IR
cutoff of HDE has been raised since its birth. As a result,
different proposals of IR cutoffs for HDE and its entropy corrected
versions \cite{FS1} have been developed.

Plan of the paper is as follows: In section II, we provide briefly
about holographical polytropic DE model and some cosmological
parameters. In IIA, we assume a particular form of Hubble parameter
and subsequently considering a correspondence between new HDE and
polytropic gas model of DE derived a new form of polytropic gas dark
energy that was further assumed to be an effective description of
dark energy in $f(T)$ gravity to study the cosmological
consequences. In section IIB we assume a particular solution for
$f(T)$ and derive solution for $H$ in the backdrop of a
correspondence between new HDE and polytropic DE. This reconstructed
$H$ has been utilized to get reconstructed effective torsion EoS and
statefinder parameters. Also, we compare the obtained results with
observational data in section IIC. In the next section III, we check
the stability of reconstructed models in all cases. We conclude the
results in section VI.

\section{New Holographic Polytropic DE in $f(T)$ Gravity}

Holographic reconstruction of modified gravity model is a very
active area of research in cosmology. Unfortunately, nature of DE is
still not known and probably that has motivated theoretical
physicists towards development of various candidates of DE and
recently geometric DE or modified gravity has been proposed as a
second approach to account for the late time acceleration of the
universe. In literature, mostly reconstructed work has been done
with polytropic EoS, family of holographic DE models, family of
Chaplygin gas, scalar field models in general relativity as well as
modified theories of gravity (in framework of $f(T)$ gravity, see
\cite{S9,S10,S11,S13,SJ,SJ3}-\cite{FS22,FS23,FS24}). However, we do
holographically reconstruction of polytropic DE and based on that we
experiment with the cosmological implications of $f(T)$ gravity.

The polytropic gas model can explain the EoS of degenerate white
dwarfs, neutron stars and also the EoS of main sequence stars.
Polytropic gas EoS is given by \cite{polyplb}
\begin{equation}\label{PDEeos}
p_{\Lambda}=K\rho_{\Lambda}^{1+\frac{1}{n}},
\end{equation}
where $K$ is a positive constant and $n$ is the polytropic index.
The important role played by polytropic EoS in astrophysics has been
emphasized in \cite{polyplb,polyepjc}. It is a simple example which
is nevertheless not too dissimilar from realistic models
\cite{polyplb}. Moreover, there are cases where a polytropic EoS is
a good approximation to reality \cite{polyplb}. From continuity
equation
\begin{equation}\label{PDE}
\rho_{\Lambda}=\left(\frac{1}{Ba^{\frac{3}{n}}-K}\right)^n.
\end{equation}
In the present work, we are considering a correspondence between
polytropic DE and new HDE with an IR cut-off proposed by
\cite{granda} with the density given by
\begin{equation}\label{NHDEdensity}
\rho_{D}=3 (\mu \dot{H}+\nu H^2).
\end{equation}

\subsection*{Statefinder and Cosmographic Parameters}

Some cosmological parameters are very important for describing the
geometry of the universe which include EoS, parameter, deceleration
parameter and statefinder etc. The physical state of a homogenous
substance can be described by EoS. This state is associated with the
matter including pressure, temperature, volume and internal energy.
It can be defined in the form $p=p(\rho, \hat{T})$, where $\rho,~p$
and $\hat{T}$ are the mass density, isotropic pressure and absolute
temperature, respectively. In cosmological context, EoS is the
relation between energy density and pressure such as $p=p(\rho)$ and
is given by
\begin{equation}\label{1.6.1a}
p=\omega\rho,
\end{equation}
where $\omega$ represents the dimensionless EoS parameter which
helps to classify different phases of the universe.

In order to differentiate different DE models on behalf of their
role in explaining the current status of the universe,  Sahni et al.
\cite{S24} proposed statefinder parameters. These are denoted by
$(r,s)$ and are defined in terms of Hubble as well as deceleration
parameters. The deceleration parameter is defined as
\begin{equation}\label{1.9.1a}
q=-\frac{\ddot{a}}{aH^2}=-\left(1+\frac{\dot{H}}{H^2}\right).
\end{equation}
The negative value of this parameter represents the accelerated
expansion of the universe due to the term $\ddot{a}>0$ (indicating
expansion with acceleration). The statefinder parameters are given
by
\begin{eqnarray}\label{1.9.2a}
r=\frac{\dddot{a}}{aH^3},\quad s=\frac{r-1}{3(q-\frac{1}{2})}.
\end{eqnarray}
These parameters possess geometrical diagnostic because of their
total dependence on the expansion factor. The most remarkable
feature of $(r,s)$ plane is that we can find the distance of a given
DE model from $\Lambda$CDM limit. This depicts the well-known
regions given as follows:\\
(i) $(r,s)=(1,0)$ shows $\Lambda$CDM limit;\\
(ii) $(r,s)=(1,1)$ describes CDM limit;\\
(iii) $r<1$ and $s>0$ constitute quintessence and phantom DE regions.\\
Moreover, $r$ can be expressed in terms of deceleration parameter as
\begin{equation}\label{1.9.5a}
r=2q^{2}+q-\frac{\dot{q}}{H}.
\end{equation}

Both $H$ and $\{r,s\}$ are categorized as cosmographic parameters.
The cosmographic parameters, being dependent on the only stringent
assumption of homogeneous and isotropic universe, marginally depend
on the choice of a given cosmological model. Secondly, cosmography
alleviates degeneracy, because it bounds only cosmological
quantities which do not strictly depend on a model. The cosmographic
set of parameters arising out of Taylor series expansion of $a(t)$
around the present epoch can be summarized as \cite{CP1,CP2}
\begin{eqnarray}\label{cosmography1}
H=\frac{1}{a}\frac{da}{dt}; ~~~~~~~~~~~~~~~~~
q=-\frac{1}{aH^2}\frac{d^2a}{dt^2}\nonumber\\
j=\frac{1}{aH^3}\frac{d^3a}{dt^3};~~~~~~~~~~~~~~~
s=\frac{1}{aH^4}\frac{d^4a}{dt^4}
\end{eqnarray}
Differentiating Friedman equation with respect to $t$ and using Eq.
\ref{cosmography1} one can write
\begin{eqnarray}
\dot{H}=-H^2 (1+q)\label{CS1}~~~~~~~~~~~~~~~~\\
\ddot{H}=H^3 (j+3q+2)\label{CS2}~~~~~~~~~~~~~\\
\dddot{H}=H^4 (s-4j-3q(q+4)-6)\label{CS3}~~~~
\end{eqnarray}
In the context of cosmological reconstruction problem, some notable
contributions are \cite{recons1,recons2,recons3}. It may be noted
that the present work is motivated by Karami and Abdolmaleki
\cite{S13}.

\subsection{With a specific choice of $H$}

We consider that the Hubble rate $H$ is given by \cite{nojiri}
\begin{equation}\label{H}
H=H_0+\frac{H_1}{t},
\end{equation}
leading to
\begin{equation}\label{a}
a(t)=C_1 e^{H_0 t} t^{H_1}.
\end{equation}
Due to this choice of Hubble parameter, the EoS takes the form
\begin{equation}\label{wlambda}
w_{\Lambda}=-1+\frac{B \left(C_1 e^{H_0 t}
t^{H_1}\right)^{\frac{1}{n}}} {-K+B \left(C_1 e^{H_0 t}
t^{H_1}\right)^{\frac{1}{n}}},
\end{equation}
and subsequently NHDE density becomes
\begin{equation}\label{NHDEt}
\rho_{D}=3 \left(-\frac{H_1 \mu
}{t^2}+\left(H_0+\frac{H_1}{t}\right)^2 \nu \right).
\end{equation}
From continuity equation we have
\begin{equation}\label{wD}
w_D=-1-\frac{\frac{2 H_1 \mu }{t^3}-\frac{2 H_1
\left(H_0+\frac{H_1}{t}\right) \nu }{t^2}}{3
\left(H_0+\frac{H_1}{t}\right) \left(-\frac{H_1 \mu
}{t^2}+\left(H_0+\frac{H_1}{t}\right)^2 \nu \right)}.
\end{equation}

\begin{figure}[ht] \begin{minipage}[b]{0.45\linewidth}
\centering\includegraphics[width=\textwidth]{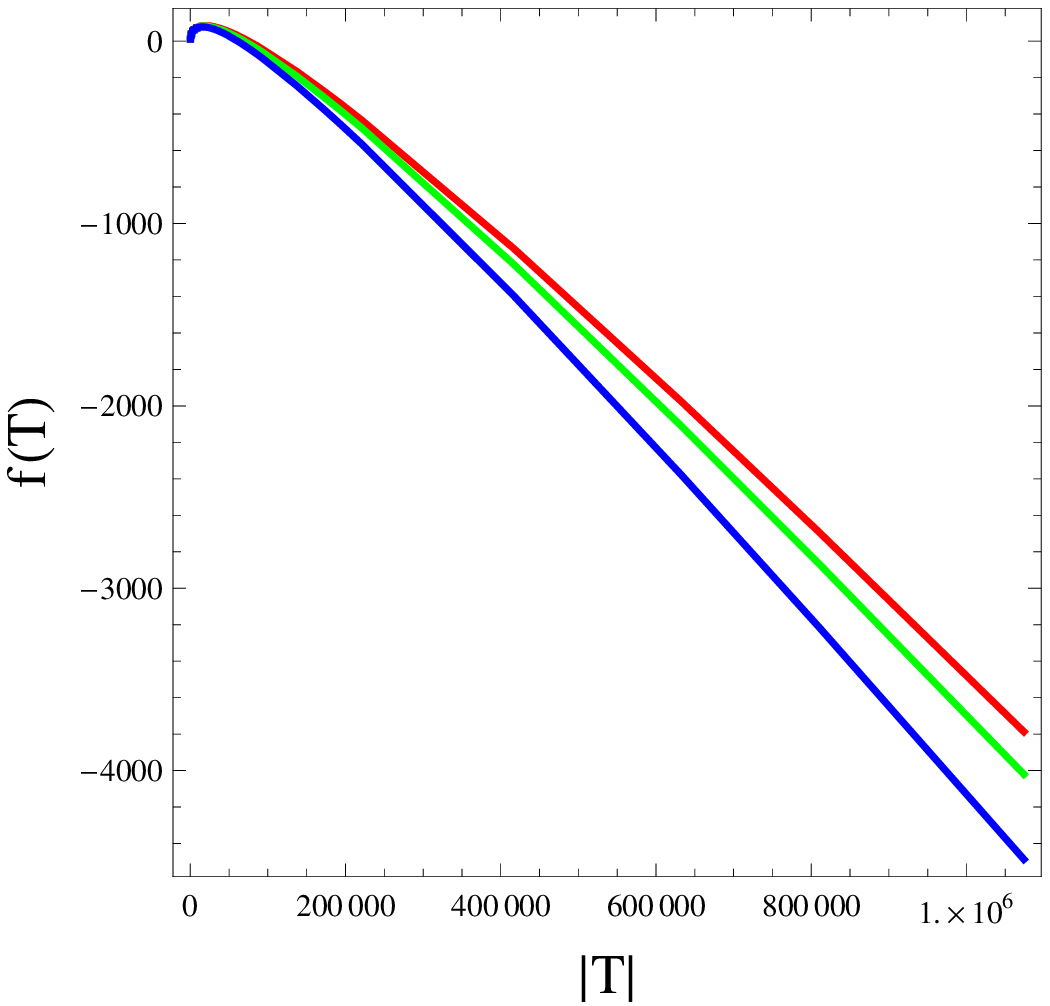}
\caption{Reconstructed $f(T)$ (Eq. (\ref{fT})) and we see that
$f(T)\rightarrow 0$ as $T\rightarrow 0$.} \label{phi2}
\end{minipage} \hspace{0.5cm} \begin{minipage}[b]{0.45\linewidth}
\centering\includegraphics[width=\textwidth]{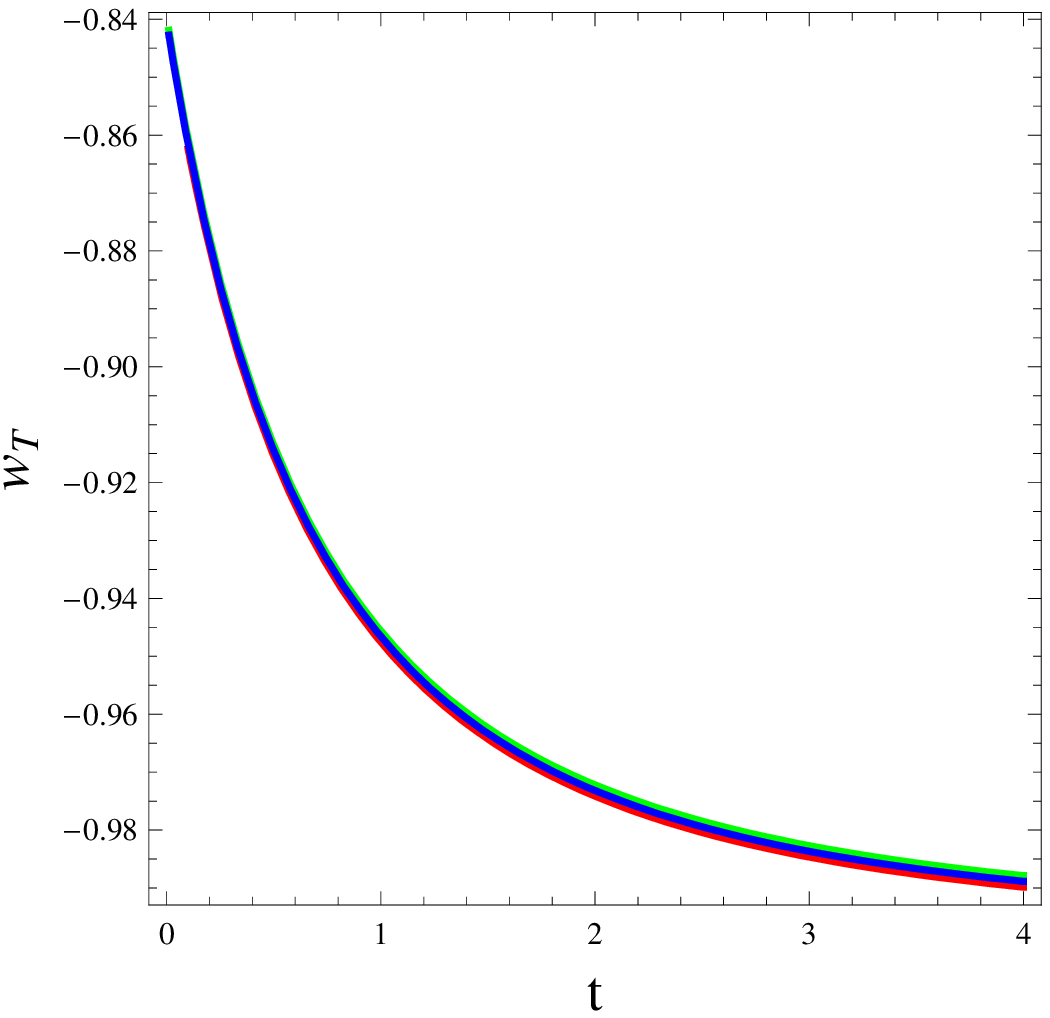}
\caption{Effective torsion EoS parameter as in Eq. (\ref{wTT}).}
\label{phia2} \end{minipage} \hspace{0.5cm}
\begin{minipage}[b]{0.45\linewidth}
\centering\includegraphics[width=\textwidth]{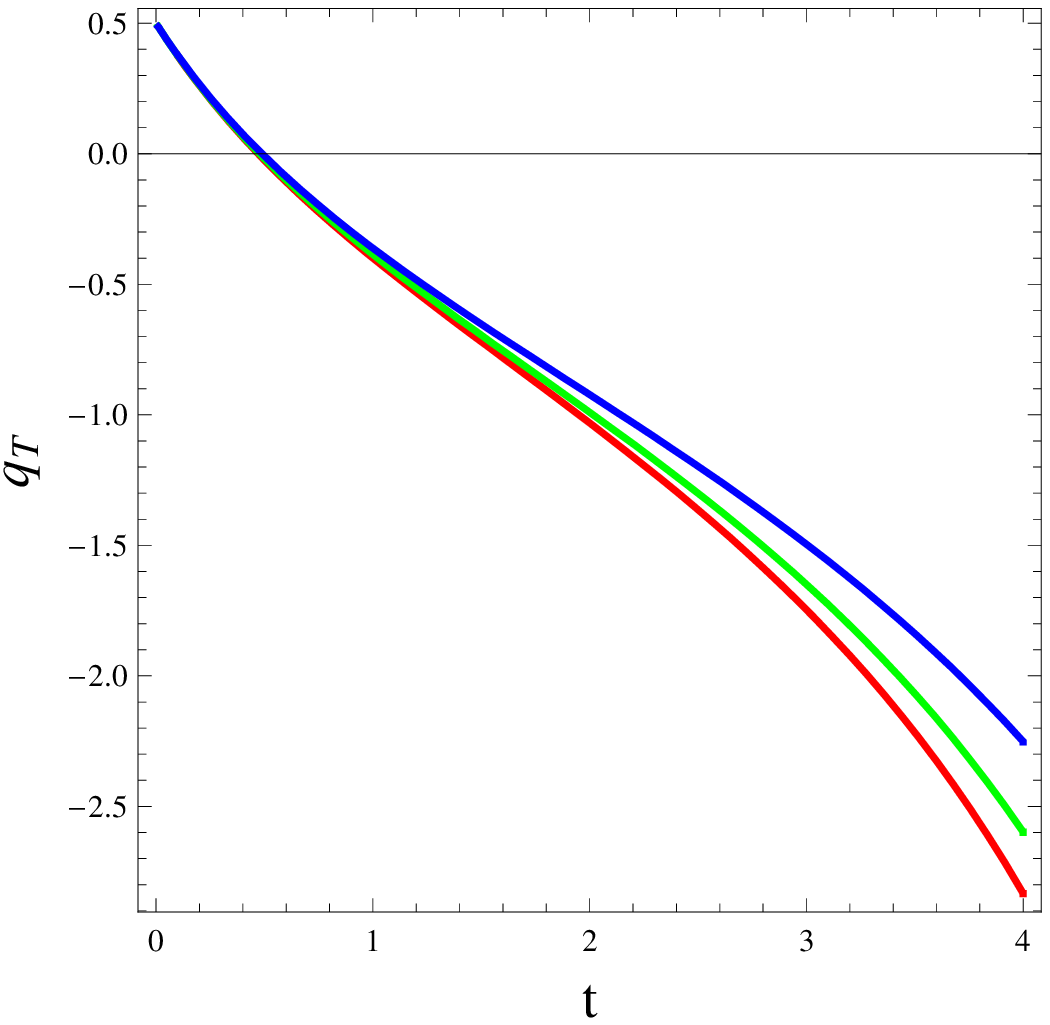}\caption{Deceleration
parameter as in Eq. (\ref{qTT}).} \label{omega2}
\end{minipage}\hspace{0.5cm}\begin{minipage}[b]{0.45\linewidth}
\centering\includegraphics[width=\textwidth]{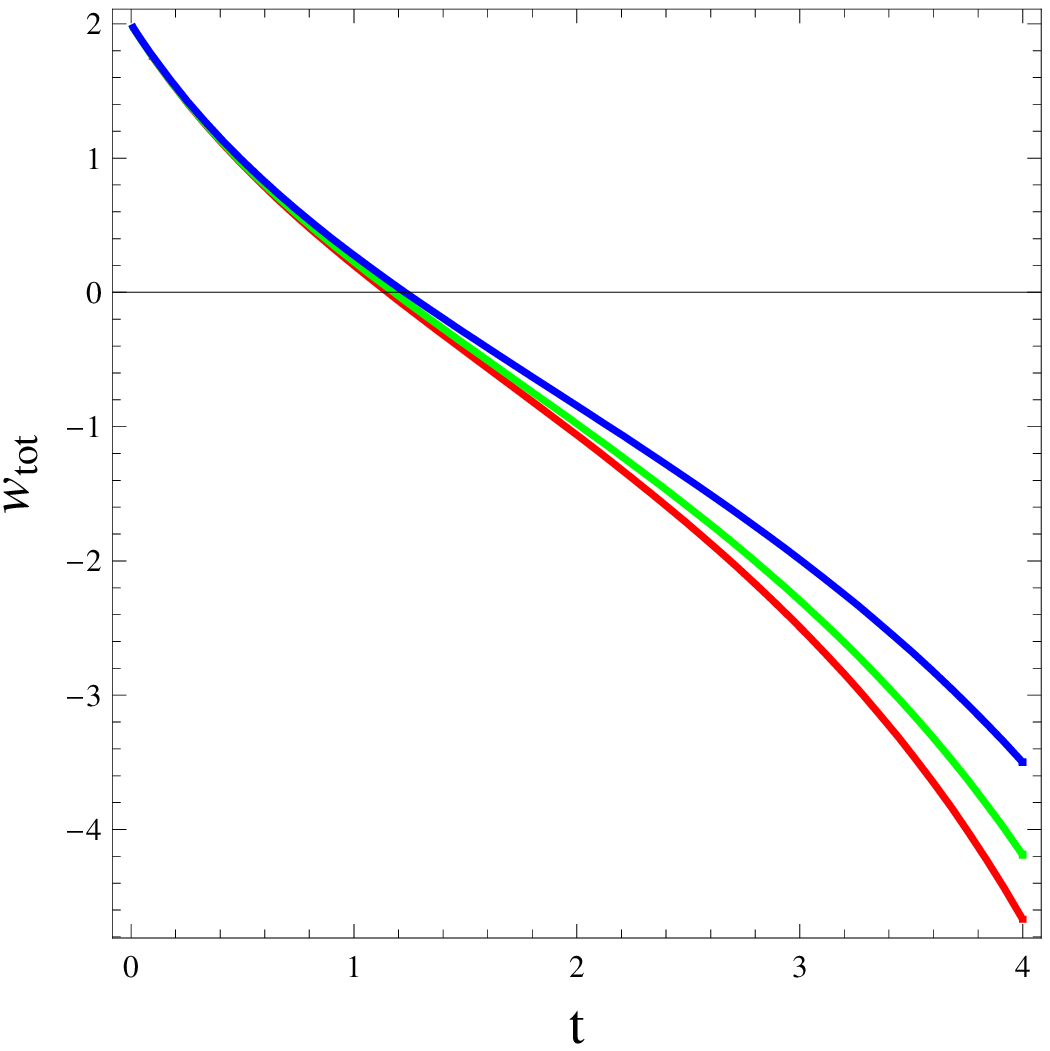}\caption{Plot
of $w_{tot}$ as in Eq. (\ref{weffTT}).}
\label{G2}\end{minipage}\end{figure}
\begin{figure}
\centering
\includegraphics[width=18pc]{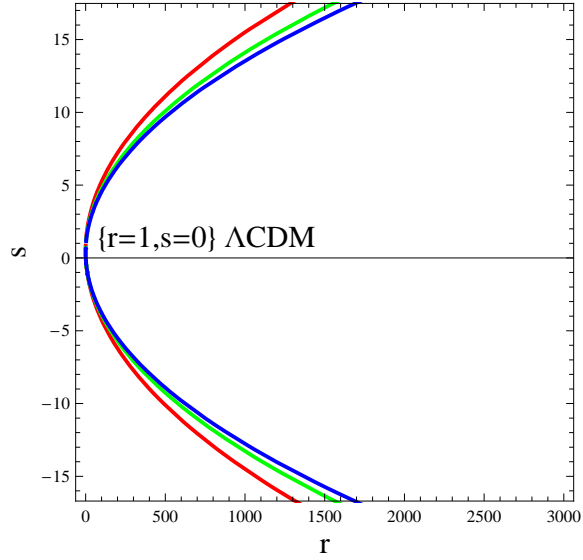}\\
\caption{Statefinder parameters for the choice of $H=H_0+\frac{H_1}{t}$.}\label{state1}
\end{figure}

Considering a correspondence between polytropic DE and new HDE i.e.
$\rho_{\Lambda}=\rho_D$ and $w_{\Lambda}=w_D$  we express $B$ and
$K$ in terms of $a$ in the following arrangement
\begin{eqnarray}
B=& \frac{2 3^{-\frac{1+n}{n}} H_1 \left(C_1 e^{H_0 t}
t^{H_1}\right)^{-1/n} (-\mu +(H_1+H_0 t) \nu )
\left(\frac{t^2}{-H_1 \mu +(H_1+H_0 t)^2 \nu }\right)^{1+\frac{1}{n}}}{t^2 (H_1+H_0 t)}\label{B} \\
K=& 3^{-\frac{1+n}{n}} \left(\frac{t^2}{-H_1 \mu +(H_1+H_0 t)^2 \nu
}\right)^{\frac{1}{n}} \left(-3+\frac{2}{H_1+H_0 t}-\frac{2 H_0 t
\nu }{-H_1 \mu +(H_1+H_0 t)^2 \nu }\right).\label{K}
\end{eqnarray}
It may be noted that $B$ and $K$ being integration constants they
are not functions of $a$. Rather it is a new arrangement arising out
of consiuderation of a correspondence between new holographic dark
energy and polytropic gas dark energy. Using Eqs. (\ref{B}) and
(\ref{K}) in Eq. (\ref{PDE}) we get the new holographic-polytropic
gas density as
\begin{equation}\label{rhoreconstruct}
\rho_{\Lambda}=3 \left(\frac{t^2}{-H_1 \mu +(H_1+H_0 t)^2 \nu }\right)^{-1}
\end{equation}
The modified Friedmann equations in the case of $f(T)$ gravity for the spatially flat FRW universe are given by
\begin{eqnarray}
H^2 &=& \frac{1}{3} \left(\rho_m + \rho_T    \right), \label{field1T} \\
2\dot{H} + 3H^2 &=& - \left(  p_m + p_T  \right), \label{field2T}
\end{eqnarray}
where
\begin{eqnarray}
\rho_T &=& \frac{1}{2} \left(2T f_T -f - T    \right), \label{rhoT} \\
p_T &=& -\frac{1}{2}\left[-8 \dot{H} f_{TT} +  \left(2T - 4\dot{H}
\right)f_T - f  + 4\dot{H}-T \right],\label{pT}  \\
T &=&-6H^2 \label{7} .
\end{eqnarray}
Here $\rho_m$ and $p_m$ are the energy density and pressure of
matter inside the universe, respectively. Also $\rho_T$ and $p_T$
are the torsion contributions to the energy density and pressure.
The energy conservation laws are given by
\begin{eqnarray}
\dot{\rho_T }+3H(\rho_T+p_T)&=& 0, \label{consv1} \\
\rho_m+3H(\rho_m+p_m)&=&0 \label{consv2} .
\end{eqnarray}
Using Eqs. (\ref{rhoT}) and (\ref{pT}), the the effective torsion
EoS parameter comes out to be
\begin{eqnarray}
w_T&=& -1+\frac{4\dot{H}(2Tf_{TT}+f_T-1)}{2Tf_T-f-T}, \label{wT}
\end{eqnarray}
Using Eqs. (\ref{field1T}), (\ref{rhoT}) and (\ref{7}) one can get
\begin{eqnarray}
\rho_m&=& \frac{1}{2}\left(f-2Tf_T\right)\label{rhomT}
\end{eqnarray}
The deceleration parameter
\begin{eqnarray}
q_T&=& 2\left(\frac{f_T-Tf_{TT}-\frac{3f}{4T}}{f_T+2Tf_{TT}}\right)\label{qT}
\end{eqnarray}
The dark torsion contribution in $f(T)$ gravity can justify the
observed acceleration of the universe without resorting to DE. This
motivates us to reconstruct an $f(T)$-gravity model according to the
new holographic-polytropic DE. Considering $\rho_T=\rho_{\Lambda}$
i.e. equating Eqs. (\ref{rhoreconstruct}) and (\ref{rhoT}) we have
the following differential equation
\begin{equation}\label{diff}
6\left(H_0+\frac{H_1}{t}\right)^2-f-\frac{t^2}{H_1}\left(H_0+\frac{H_1}{t}\right)
\frac{df}{dt}=6\left(\frac{t^2}{-\mu H_1+(H_0 t+H_1)^2
\nu}\right)^{-1}.
\end{equation}
Solving (\ref{diff}), we obtain reconstructed $f$ in terms of cosmic
time $t$
\begin{equation}
\begin{array}{c}\label{ft}
f(t)=\frac{1}{H_1 t^2}\left[H_1 \left\{H_0 t (C_2 t-12 \mu )+H_1 (C_2 t-6 \mu
+6 H_0 t (-1+\nu ))+6 H_1^2 (-1+\nu )\right\}\right.\\
\left.+12 H_0 t (H_1+H_0 t) \mu  \ln(\frac{H_1}{t}+H_0)\right].
\end{array}
\end{equation}
Considering $H=(-\frac{T}{6})^{1/2}$, we have
\begin{equation}
t=\frac{6H_1}{-6H_0+\sqrt{-6T}},
\end{equation}
that lead to re-expressing $f$ of (\ref{ft}) as a function of $T$ as
\begin{equation}\label{fT}
f(T)=\frac{ C_2H_1 \sqrt{-6T}+6 \left(6 H_0^2 \mu - H_0 H_1
\sqrt{-6T} (-1+\nu )+T (H_1+\mu -H_1 \nu )\right)+6 H_0 \sqrt{-6T}
\mu  \ln\left[-\frac{T}{6}\right]}{6 H_1}.
\end{equation}
Subsequently using (\ref{fT}) in (\ref{wT}) and (\ref{qT}), we get
the effective torsion EoS and deceleration parameters as
\begin{equation}\label{wTT}
w_T=-1+\frac{\left(-6 H_0+\sqrt{-6T}  \right)^2 \left(6 H_0 \mu
+\sqrt{-6T}   (-\mu +H_1 \nu )\right)}{9 \sqrt{-6T} H_1 \left(-6
H_0^2 \mu +2 \sqrt{-6T} H_0  \mu +T (\mu -H_1 \nu )\right)},
\end{equation}
and
\begin{equation}\label{qTT}
q_T=\frac{- \left(-18 H_0^2+4\sqrt{-6T} H_0 +T\right) \mu +H_1 T
\left(-1+ \nu \right)} {-2  \left(H_0\sqrt{-6T}+T\right) \mu +2 H_1
T \left(-1+ \nu \right)},
\end{equation}
Using Eq. (\ref{fT}) in (\ref{rhomT}) density of the dark matter inside the universe becomes
\begin{equation}\label{rhomTT}
\rho_m=\frac{\left(6 H_0^2-2\sqrt{-6T} H_0 -T\right) \mu +H_1 T
(-1+\nu )}{2 H_1}.
\end{equation}
In the case of pressureless dust matter, $p_m=0$, we obtain
\begin{equation}\label{hdotT}
\dot{H}=-\frac{1}{2}\left(\frac{\rho_m}{f_T+2Tf_{TT}}\right).
\end{equation}
Using Eqs. (\ref{fT}) and (\ref{rhomTT}) in (\ref{hdotT}) we get
\begin{equation}\label{hdotTT}
\dot{H}=\frac{ \sqrt{\frac{-3T}{2}} \left(2 H_0 \left(-3
H_0+\sqrt{-6T}\right) \mu +T (H_1(1-\nu)+\mu )\right)}{2 \left(-6
H_0 \mu +\sqrt{-6T} (H_1(1-\nu)+\mu)\right)}.
\end{equation}
Defining the effective energy-density and pressure as
$\rho_{tot}=\rho_T+\rho_m$ and $p_{tot}=p_T~(p_m=0)$ the effective
EoS $w_{tot}=p_{tot}/\rho_{tot}$ becomes (using Eq. (\ref{hdotTT}))
\begin{equation}\label{weffTT}
w_{tot}=-1-\frac{2 \dot{H}}{3H^2}=-1-\frac{3 \sqrt{6} \left(2 H_0
\left(-3 H_0+ \sqrt{-6T}\right) \mu +T (H_1(1-\nu)+\mu )\right)}{6
H_0 \sqrt{-T} \mu +\sqrt{6} T (H_1(1-\nu)+\mu )}.
\end{equation}

The statefinder parameters are given by
\begin{eqnarray}
r &=& q+2q^2+\frac{\dot{q}}{H}\label{r} \\
s &=& \frac{r-1}{3\left(q-\frac{1}{2}\right)}. \label{s}
\end{eqnarray}
In the current framework Eq.(\ref{r}) and (\ref{s}) take the form
\begin{equation}\label{rT}
\begin{array}{c}
r=-\frac{1}{H_1 T^2 \left(-6 H_0 \mu +\sqrt{-6T} (H_1+\mu -H_1 \nu )\right)^2}\times \\
3\left(-3 H_0 \left(36
\sqrt{6} H_0^4 \sqrt{-T}+36 \sqrt{6} H_0^2 (-T)^{3/2}+\sqrt{6} (-T)^{5/2}+144 H_0^3
T-24 H_0 T^2\right) \mu ^2-\right.\\
\left.H_1^2
T^2 \left(-54 H_0^2+13 H_0\sqrt{-6T}+4 T\right) \mu  (-1+\nu )+\right.\\
\left.2 H_1^3 T^3 (-1+\nu )^2+H_1 T \mu  \left(2 T^2 \mu
-6 H_0^2 T (29 \mu +12 (-1+\nu ))+108 H_0^4 (-2+3 \mu +2 \nu )+\right.\right.\\
\left.\left.H_0\sqrt{-6T} T (-3+13 \mu +3 \nu )-18 \sqrt{6} H_0^3
\sqrt{-T} (-5+9 \mu +5 \nu )\right)\right).
\end{array}
\end{equation}
\begin{equation}\label{sT}
\begin{array}{c}
s=\frac{1}{3 H_1 T \left(6 H_0 \mu +\sqrt{6} \sqrt{-T} (-H_1-\mu
+H_1 \nu )\right)}\times\\
\left[36 H_0^3 \mu +\sqrt{6} (-1+3 H_1) \sqrt{-T} T (-H_1-\mu +H_1 \nu )+\right.\\
\left.18 H_0 T (-\mu +H_1 (-1+2
\mu +\nu ))+6 \sqrt{6} H_0^2 \sqrt{-T} (-3 \mu +H_1 (-2+3 \mu +2 \nu ))\right]
\end{array}.
\end{equation}
It may be noted that in the present and subsequent figures red,
green and blue lines correspond to $n=6,~8$ and $10$ respectively.
Figure \textbf{2} shows the evolution of the effective torsion EoS
parameter $w_T$ as a function of $t$. In this case $w_T>-1$ and it
is running close to $-1$, but it is not crossing $-1$ boundary. This
indicates ``quintessence" behavior. In later time
$-6H_0+\sqrt{-6T}\rightarrow 0$ (see Eq. (\ref{wTT})) and as a
consequence $w_T\rightarrow -1$. A clear transition from $q>0$ to
$q<0$ is apparent at $t\approx 0.5$ in Figure \textbf{3}. This
indicates transition from decelerated to accelerated phase of the
universe. In Figure \textbf{4}, it is observed that $w_{tot}$
behaves differently from $w_{eff}$. The $w_{tot}$ transits from
$>-1$ i.e. quintessence to $<-1$ i.e. phantom at $t\approx 1$.
Statefinder parameters as obtained in Eqs. (\ref{rT}) and (\ref{sT})
are plotted in Figure \textbf{5} and it is observed that the fixed
point $\{r=1,s=0\}_{\Lambda CDM}$ is attainable and the $\{r-s\}$
trajectory goes beyond the $\Lambda$CDM. It is palpable that for
finite $r$, we have $s\rightarrow -\infty$. This indicates that the
holographic-polytropic $f(T)$ gravity interpolates between dust and
$\Lambda$CDM phase of the universe. In this framework, the
cosmographic parameter $j$ (jerk) comes out to be
\begin{equation}\label{jerkwithout}
j=-\frac{7}{2}+\frac{2 H_1}{(H_1+H_0 t)^3}+\frac{9 H_0 t \mu }{2 (H_1+H_0 t) (H_1+H_0 t+\mu
-(H_1+H_0 t) \nu )}
\end{equation}

\subsection{With specific form of $f$ and without any assumption
about $H$}

\subsection*{Power-law model of Bengochea and Ferraro}

In this section, we are not assuming any form of $H$ or $a$. Rather
we assume $f$ as the power-law model of Bengochea and Ferraro
\cite{BF}
\begin{equation}\label{fBF}
f(T)=\alpha (-T)^b
\end{equation}
where $\alpha$ and $b$ are the two model parameters. Considering $\rho_{\Lambda}=\rho_D$ we have
the following differential equation
\begin{equation}\label{diffHsqr}
\frac{\mu a}{2}\left(\frac{dH^2}{da}\right)+\nu H^2=\frac{1}{3}\left(Ba^{1/n}-K\right)^{-n}
\end{equation}
solving which we get
\begin{equation}\label{Hsqr}
H^2=a^{-\frac{2 \nu }{\mu }} C_1+\frac{\left(1-\frac{a^{\frac{1}{n}} B}{K}\right)^n \left(a^{\frac{1}{n}}
B-K\right)^{-n} 2F1\left[\frac{2
n \nu }{\mu },n,1+\frac{2 n \nu }{\mu },\frac{a^{\frac{1}{n}} B}{K}\right]}{3 \nu }
\end{equation}
that leads to
\begin{equation}\label{hdota}
\dot{H}=\frac{-3 a^{-\frac{2 \nu }{\mu }} C_1 \nu +\left(a^{\frac{1}{n}} B-K\right)^{-n} \left(1-
\left(1-\frac{a^{\frac{1}{n}} B}{K}\right)^n
2F1\left[\frac{2 n \nu }{\mu },n,1+\frac{2 n \nu }{\mu },\frac{a^{\frac{1}{n}} B}{K}\right]\right)}{3 \mu }
\end{equation}
Therefore, using $T=-6H^2$ in Eq. (\ref{fBF}) and thereafter using (\ref{rhomT}) we have the dark
matter density of the universe as a function of $a$ as
\begin{equation}\label{rhomBF}
\rho_m=\frac{1}{2} (1-2 b) \alpha  \left(6 a^{-\frac{2 \nu }{\mu }} C_1+\frac{2 \left(1-\frac{a^{\frac{1}{n}} B}
{K}\right)^n \left(a^{\frac{1}{n}}
B-K\right)^{-n} 2F1\left[\frac{2 n \nu }{\mu },n,1+\frac{2 n \nu }{\mu },
\frac{a^{\frac{1}{n}} B}{K}\right]}{\nu }\right)^b
\end{equation}
Using Eq. (\ref{rhomBF}) in (\ref{hdotT}) we have for the present choice of $f(T)$
\begin{equation}\label{hdot28}
\dot{H}=-\frac{3 a^{-\frac{2 \nu }{\mu }} C_1+\frac{\left(1-\frac{a^{\frac{1}{n}} B}{K}\right)^n
\left(a^{\frac{1}{n}} B-K\right)^{-n} 2F1\left[\frac{2
n \nu }{\mu },n,1+\frac{2 n \nu }{\mu },\frac{a^{\frac{1}{n}} B}{K}\right]}{\nu }}{2 b}
\end{equation}
As we are considering new holographic polytropic dark energy in $f(T)$ gravity, we can consider
equality of Eqs. (\ref{hdot28}) and (\ref{hdota}) from which we can express the integration constant $C_1$ as
 \begin{equation}\label{C1}
C_1=\frac{a^{\frac{2 \nu }{\mu }} \left(a^{\frac{1}{n}} B-K\right)^{-n} \left(2 b \nu +
\left(1-\frac{a^{\frac{1}{n}} B}{K}\right)^n (3 \mu
-2 b \nu ) 2F1\left[\frac{2 n \nu }{\mu },n,1+\frac{2 n \nu }{\mu },
\frac{a^{\frac{1}{n}} B}{K}\right]\right)}{3 \nu  (-3 \mu
+2 b \nu )}
\end{equation}
As Eq. (\ref{C1}) is used in (\ref{Hsqr}) the $H^2$ reduces to
\begin{equation}\label{Hsqrreduced}
H^2=-\frac{2 b \left(a^{\frac{1}{n}} B-K\right)^{-n}}{9 \mu -6 b \nu }
\end{equation}
and hence
\begin{equation}\label{Hdotreduced}
\dot{H}=\frac{a^{\frac{1}{n}} b B \left(a^{\frac{1}{n}} B-K\right)^{-1-n}}{9 \mu -6 b \nu }
\end{equation}
Subsequently, effective torsion EoS and deceleration parameters become
\begin{equation}\label{wTreduced}
w_T=-1+\frac{a^{\frac{1}{n}} b B \left(4+\alpha  \left(\frac{b \left(a^{\frac{1}{n}} B-K\right)^{-n}}
{-3 \mu +2 b \nu }\right)^b \left(-2^{3+2
b} b+\left(a^{\frac{1}{n}} B-K\right)^n \left(-3 4^b \mu +2^{1+2 b} b \nu \right)\right)\right)}
{3 \left(a^{\frac{1}{n}} B-K\right) \left(4 b+\left(a^{\frac{1}{n}}
B-K\right)^n \alpha  \left(\frac{b \left(a^{\frac{1}{n}} B-K\right)^{-n}}{-3 \mu +2 b \nu }\right)^b
 \left(3 4^b \mu +2^{1+2 b} b (-3 \mu +(-1+2
b) \nu )\right)\right)}
\end{equation}
\begin{equation}\label{qTreduced}
q_T=\frac{4-\frac{3}{b}+\frac{16 b \left(a^{\frac{1}{n}} B-K\right)^{-n}}{-3 \mu +2 b \nu }}{2 (-1+2 b)}
\end{equation}
\begin{equation}\label{weffreduced}
w_{tot}=-1+\frac{4 a^{\frac{1}{n}} B+\left(a^{\frac{1}{n}} \left(3
4^b-7 4^b b+2^{1+2 b} b^2\right) B-3 4^b K+3 2^{1+2 b} b K\right)
\alpha \left(\frac{b \left(a^{\frac{1}{n}} B-K\right)^{-n}}{-3 \mu
+2 b \nu }\right)^{-1+b}}{12 \left(a^{\frac{1}{n}} B-K\right)}
\end{equation}
In this framework Eqs.(\ref{r}) and (\ref{s}) take the form
\begin{eqnarray}
r=\frac{\left(4-\frac{3}{b}+\frac{16 b \left(a^{\frac{1}{n}} B-K\right)^{-n}}{-3 \mu +2 b \nu }\right)
\left(3-\frac{3}{b}+b \left(2+\frac{16
\left(a^{\frac{1}{n}} B-K\right)^{-n}}{-3 \mu +2 b \nu }\right)\right)}{2 (1-2 b)^2}\label{rTreduced}\\
s=\frac{2-\frac{3}{b}+b \left(4+\frac{16 \left(a^{\frac{1}{n}} B-K\right)^{-n}}{-3 \mu +2 b \nu }\right)}
{3 (-1+2 b)}~~~~~~~~~~~~~~~~~~~~~~~~~~~~~~~~~~~~~~~~\label{sTreduced}
\end{eqnarray}
and the other cosmographic parameter $j$ (jerk parameter) (using Eq. (\ref{CS3})) takes the form

\begin{equation}\label{jerkwith}
j=\frac{1}{2} \left(-4+\frac{a^{\frac{1}{n}} B \left(K+a^{\frac{1}{n}} B n\right)}
{\left(-a^{\frac{1}{n}} B+K\right)^2 n}+\frac{3 \left(-4+\frac{3}{b}+\frac{16
b \left(a^{\frac{1}{n}} B-K\right)^{-n}}{3 \mu -2 b \nu }\right)}{-1+2 b}\right)
\end{equation}
\begin{figure}[ht] \begin{minipage}[b]{0.45\linewidth}
\centering\includegraphics[width=\textwidth]{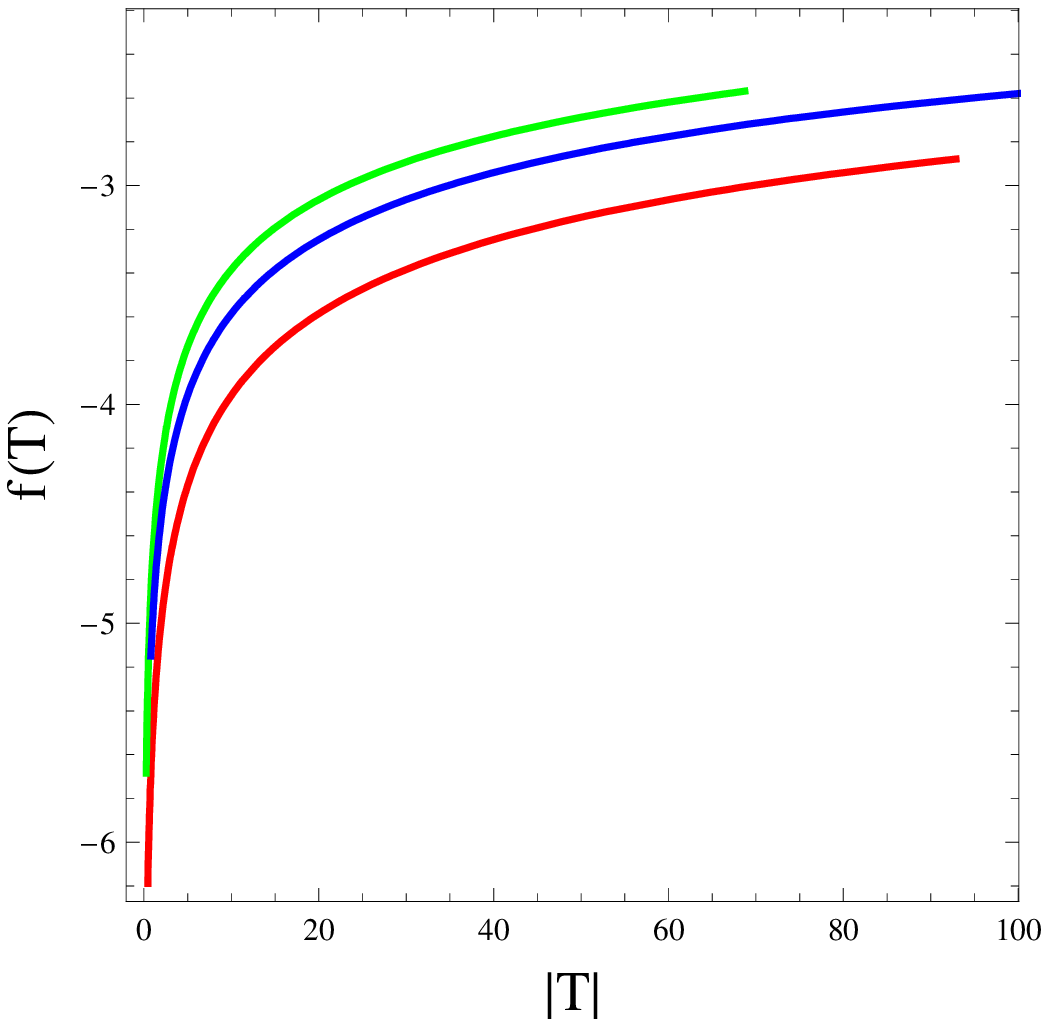} \caption{Plot
of $f(T)$ based on reconstructed $H$.} \label{phi2} \end{minipage}
\hspace{0.5cm} \begin{minipage}[b]{0.45\linewidth}
\centering\includegraphics[width=\textwidth]{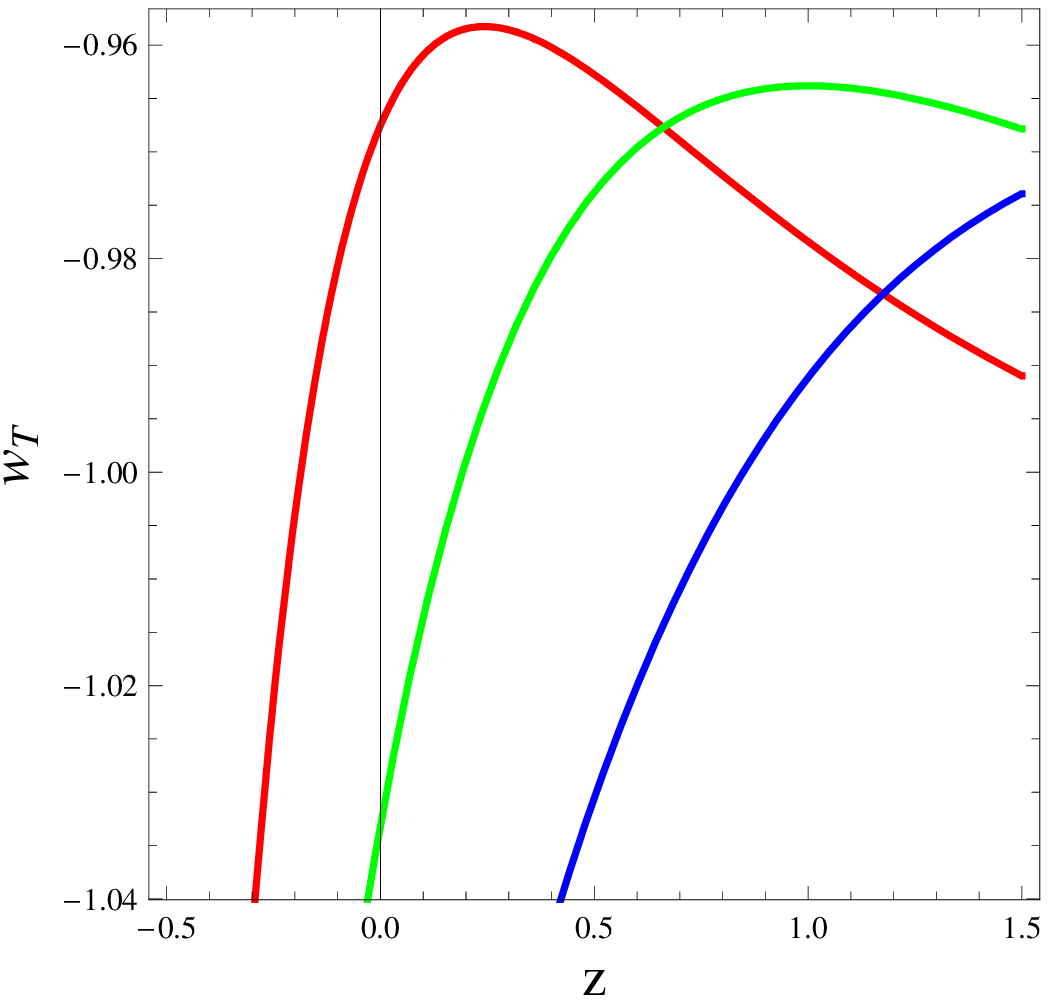}
\caption{Effective torsion EoS parameter for the Bengochea and
Ferraro model.} \label{phia2} \end{minipage}
\hspace{0.5cm}
\begin{minipage}[b]{0.45\linewidth}
\centering\includegraphics[width=\textwidth]{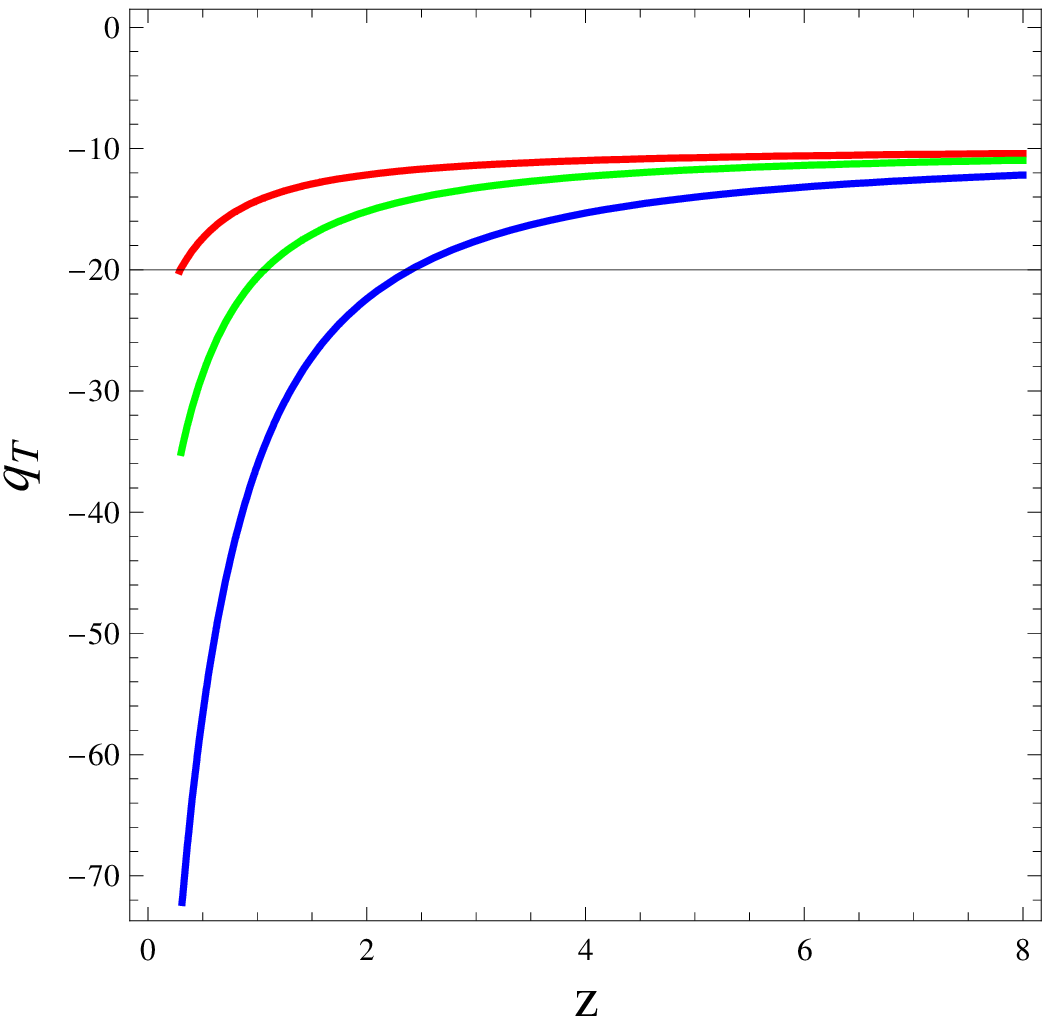}\caption{Deceleration
parameter $q_T$ for the Bengochea and Ferraro model.}\label{phia2}
\end{minipage} \hspace{0.5cm} \begin{minipage}[b]{0.45\linewidth}
\centering\includegraphics[width=\textwidth]{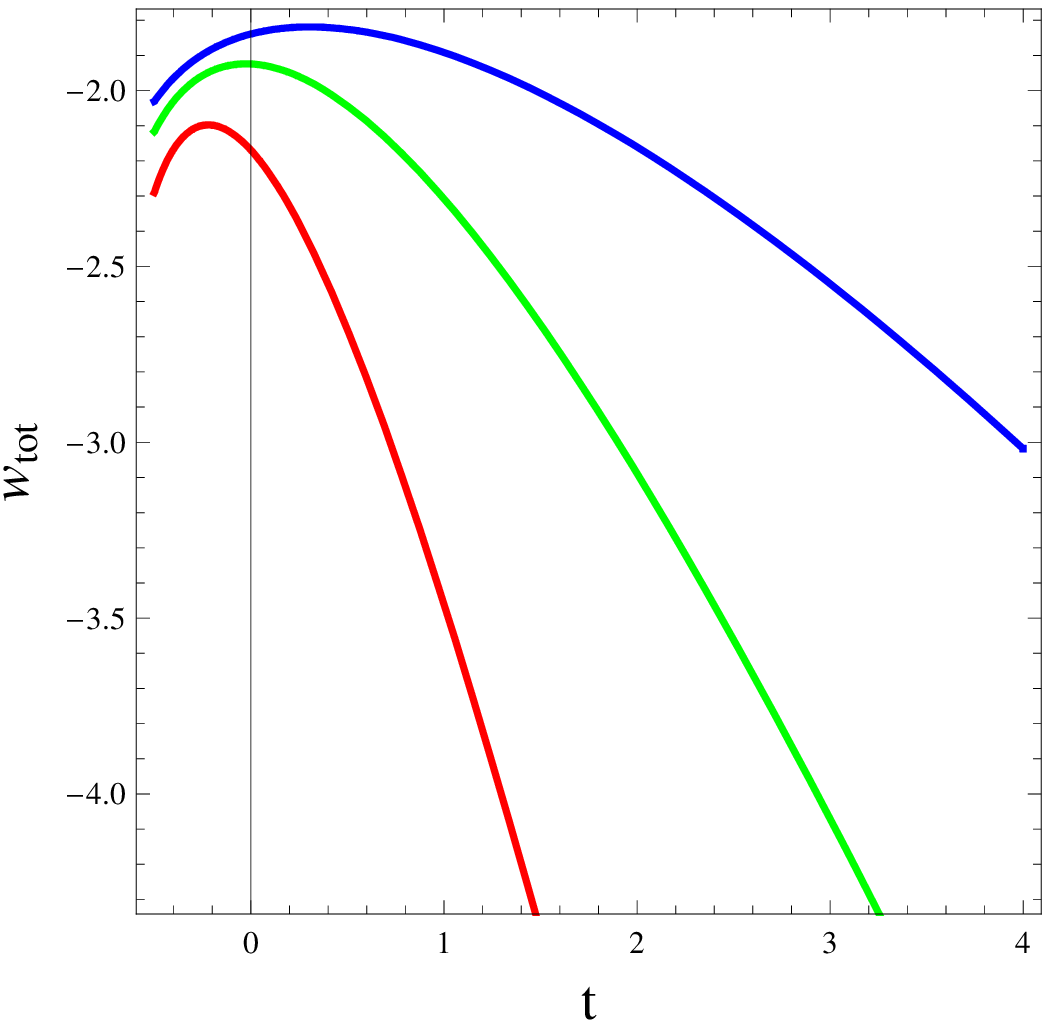}\caption{Plot
of $w_{tot}$ as in Eq. (\ref{weffreduced}).} \label{omega2}
\end{minipage}\end{figure}

\begin{figure}
\centering
\includegraphics[width=18pc]{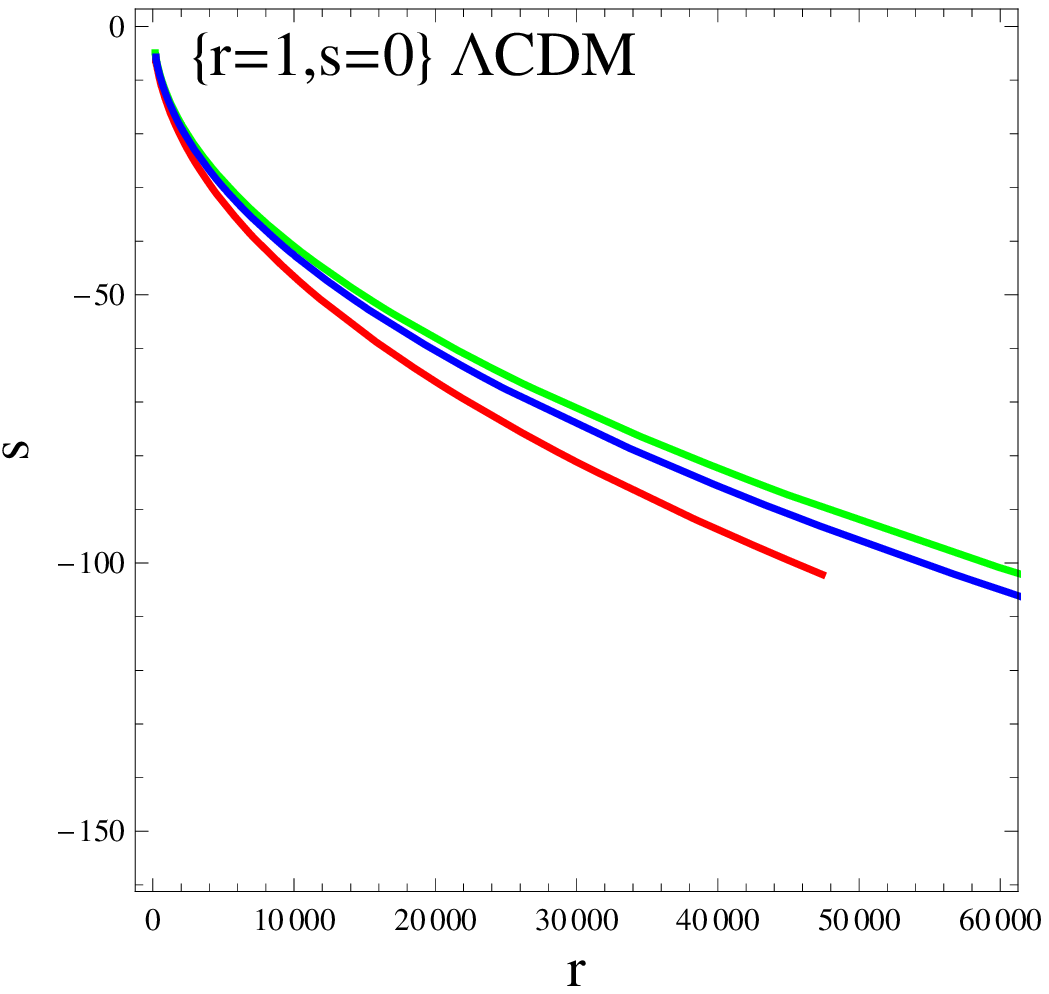}\\
\caption{Statefinder parameters for the choice of $f(T)=\alpha (-T)^b$. }\label{state1}
\end{figure}

In Figure 6, $f(T)$ is plotted against $T$ and it is observed that $f(T)\rightarrow -\infty$
as $T\rightarrow 0$. The effective torsion parameter is plotted in
Figure 7 and it is palpable that $w_{T}<-1$ i.e. behaves like
phantom. The deceleration parameter plotted in Figure 8 shows an
ever accelerating universe. The $w_{tot}<-1$ i.e. behaves like
phantom as seen in Figure 9. The statefinders as obtained in Eqs.
(\ref{rTreduced}) and (\ref{sTreduced}) are plotted in Figure 10 and
the $\{r-s\}$ trajectory attains the $\Lambda$CDM point i.e.
$\{r=1,s=0\}$. However, unlike the previous model the dust phase is
not apparently attained by the statefinder trajectory.

\subsection*{Exponential model}

We consider exponential $f(T)$ gravity \cite{kaju1}
\begin{equation}\label{expF}
f(T)=\delta \exp (\xi T)
\end{equation}
Subsequently, using $T=-6H^2$ in Eq. (\ref{expF}), where $H^2$ is as
obtained in Eq.(\ref{Hsqr}), and thereafter using (\ref{rhomT}) we
have the dark matter density of the universe as a function of $a$
for present choice of $f$ as
\begin{eqnarray}\label{exporhom}
\rho_m=\frac{1}{2 \nu}a^{-\frac{2 \nu }{\mu }} e^{\left(-6 a^{-\frac{2 \nu }{\mu }}
C_1 \xi -\frac{2 \left(1-\frac{a^{\frac{1}{n}} B}{K}\right)^n \left(a^{\frac{1}{n}}
B-K\right)^{-n} \xi  2F1\left[\frac{2 n \nu }{\mu },n,1+\frac{2 n \nu }{\mu },
\frac{a^{\frac{1}{n}} B}{K}\right]}{\nu }\right)} \left(a^{\frac{1}{n}} B-K\right)^{-n}\times\nonumber \\
\delta
 \left(\left(a^{\frac{1}{n}} B-K\right)^n \nu  \left(a^{\frac{2 \nu }{\mu }}+12 C_1 \xi
 \right)+4 a^{\frac{2 \nu }{\mu }} \left(1-\frac{a^{\frac{1}{n}} B}{K}\right)^n \xi
 2F1\left[\frac{2 n \nu }{\mu },n,1+\frac{2 n \nu }{\mu },\frac{a^{\frac{1}{n}} B}{K}\right]\right)~~~~
\end{eqnarray}
Using Eq. (\ref{exporhom}) in (\ref{hdotT}) we have for the present choice of $f(T)$
\begin{eqnarray}\label{expHdot}
\dot{H}=\frac{\left(a^{\frac{1}{n}} B-K\right)^n \nu
\left(a^{\frac{2 \nu }{\mu }}+ 12 C_1 \xi \right)+4 a^{\frac{2 \nu
}{\mu }} \left(1-\frac{a^{\frac{1}{n}} B}{K}\right)^n \xi
2F1\left[\frac{2 n \nu }{\mu },n,1+\frac{2 n \nu }{\mu
},\frac{a^{\frac{1}{n}} B}{K}\right]}{4 \xi
\left(-\left(a^{\frac{1}{n}} B-K\right)^n \nu \left(a^{\frac{2 \nu
}{\mu }}-12 C_1 \xi \right)+4 a^{\frac{2 \nu }{\mu }}
\left(1-\frac{a^{\frac{1}{n}} B}{K}\right)^n \xi  2F1\left[\frac{2 n
\nu }{\mu },n,1+\frac{2 n \nu }{\mu },\frac{a^{\frac{1}{n}}
B}{K}\right]\right)}
\end{eqnarray}
Considering equality of Eqs. (\ref{hdot28}) and (\ref{expHdot}) we can express $C_1$ as
\begin{eqnarray}\label{expC1}
C_1=\frac{1}{24 \nu ^2 \xi ^2}\left(a^{\frac{1}{n}} B-K\right)^{-2
n} \left(\left(a^{\frac{4 \nu }{\mu }} \left(a^{\frac{1}{n}}
B-K\right)^{2 n} \nu ^2 \xi ^2 \left(\left(a^{\frac{1}{n}} B-K\right)^{2 n}\right.\right.\right.\times\nonumber\\
\left.\left.\left. \left(9 \mu ^2-18 \mu  \nu +\nu ^2\right)-8
\left(a^{\frac{1}{n}} B-K\right)^n (3 \mu +\nu ) \xi +16 \xi
^2\right)\right)^{1/2}+ a^{\frac{2 \nu }{\mu }}
\left(a^{\frac{1}{n}} B-K\right)^n \nu  \xi
\right.\times\nonumber\\
\left. \left(\left(a^{\frac{1}{n}} B-K\right)^n (-3 \mu +\nu )+4 \xi
-8 \left(1-\frac{a^{\frac{1}{n}} B}{K}\right)^n \xi 2F1\left[\frac{2
n \nu }{\mu },n,1+\frac{2 n \nu }{\mu },\frac{a^{\frac{1}{n}}
B}{K}\right]\right)\right)
\end{eqnarray}
that finally leads to
\begin{equation}
\begin{array}{c}\label{expC1hdot}
\dot{H}=\left(a^{\frac{1}{n}} B \left(a^{\frac{1}{n}}
B-K\right)^{-1-n} \left(a^{\frac{2 \nu }{\mu }} \left(a^{\frac{1}{n}}
B-K\right)^n \nu  \left(\left(a^{\frac{1}{n}} B-K\right)^n (3 \mu +\nu )-4 \xi \right) \xi\right.\right.\\
\left.\left. -\sqrt{a^{\frac{4 \nu }{\mu }} \left(a^{\frac{1}{n}}
B-K\right)^{2 n} \nu ^2 \xi ^2 \left(\left(a^{\frac{1}{n}} B-K\right)^{2 n}
\left(9 \mu ^2-18 \mu  \nu +\nu ^2\right)-8 \left(a^{\frac{1}{n}}
B-K\right)^n (3 \mu +\nu ) \xi +16 \xi ^2\right)}\right)\right)\\
\left(12 \nu  \sqrt{a^{\frac{4 \nu }{\mu }} \left(a^{\frac{1}{n}}
B-K\right)^{2 n} \nu ^2 \xi ^2 \left(\left(a^{\frac{1}{n}} B-K\right)^{2 n}
\left(9 \mu ^2-18 \mu  \nu +\nu ^2\right)-8 \left(a^{\frac{1}{n}} B-K\right)^n
(3 \mu +\nu ) \xi +16 \xi ^2\right)}\right)^{-1}
\end{array}
\end{equation}

\begin{figure}[ht] \begin{minipage}[b]{0.45\linewidth}
\centering\includegraphics[width=\textwidth]{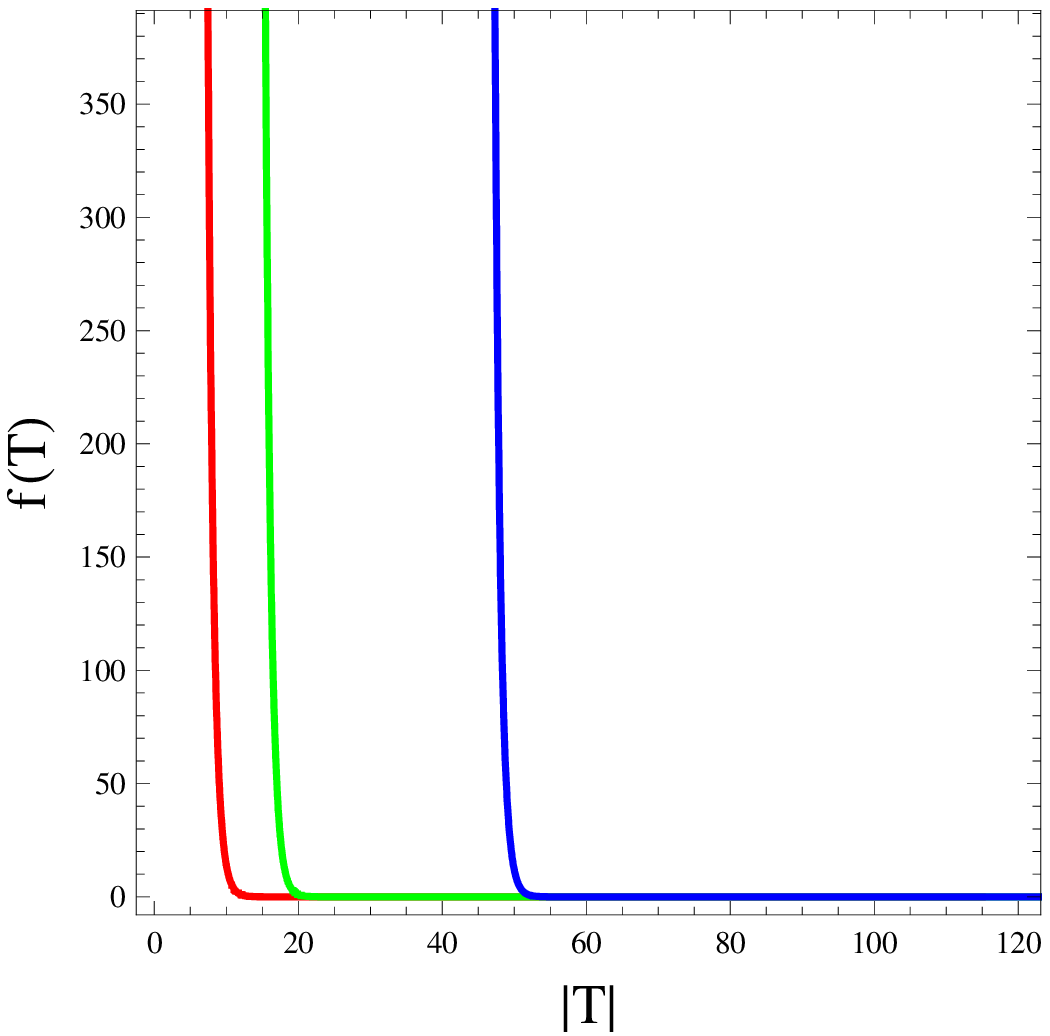}
\caption{Reconstructed $f(T)$ for the exponential model and we see that
$f(T)$ becomes a decreasing function of $T$.} \label{11}
\end{minipage} \hspace{0.5cm} \begin{minipage}[b]{0.45\linewidth}
\centering\includegraphics[width=\textwidth]{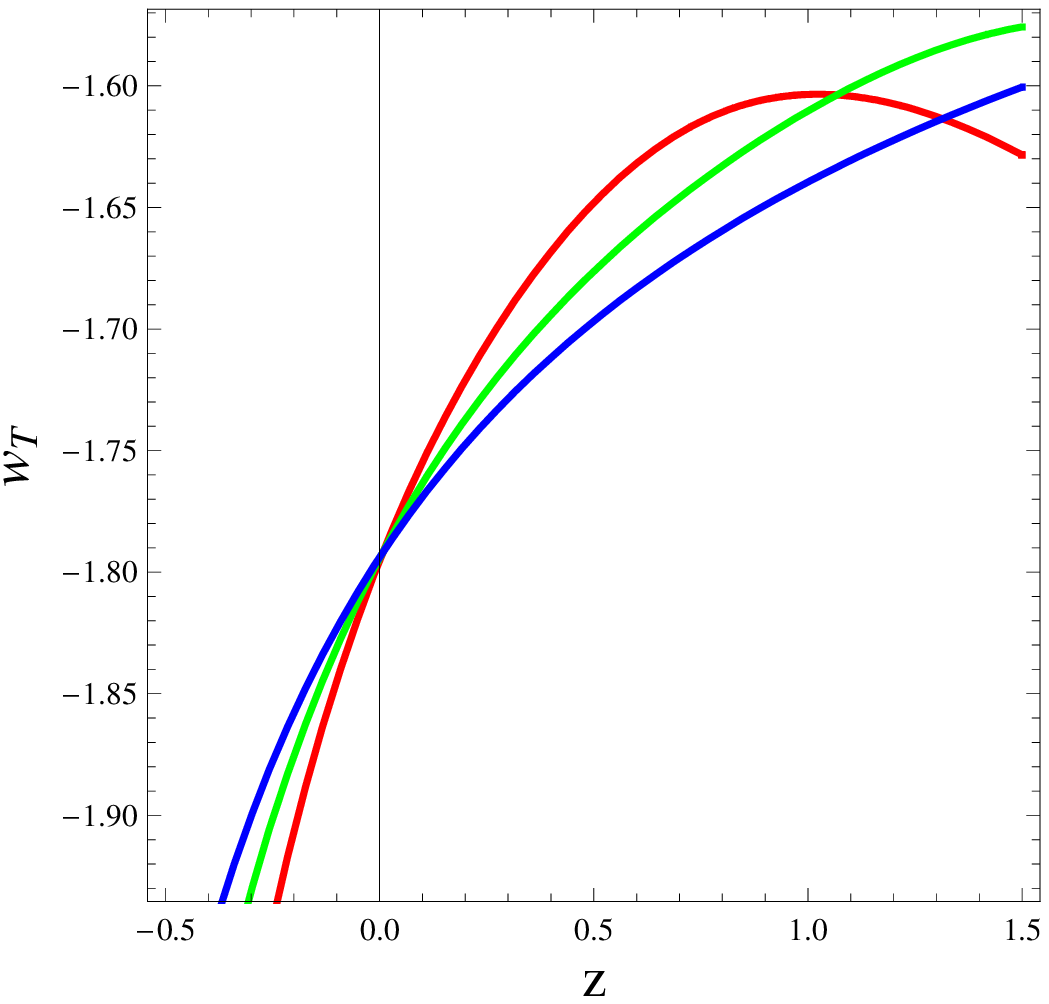}
\caption{Effective torsion EoS parameter based on reconstructed $\dot{H}$ in Eq. (\ref{expC1hdot}).}
\label{12} \end{minipage} \hspace{0.5cm}
\begin{minipage}[b]{0.45\linewidth}
\centering\includegraphics[width=\textwidth]{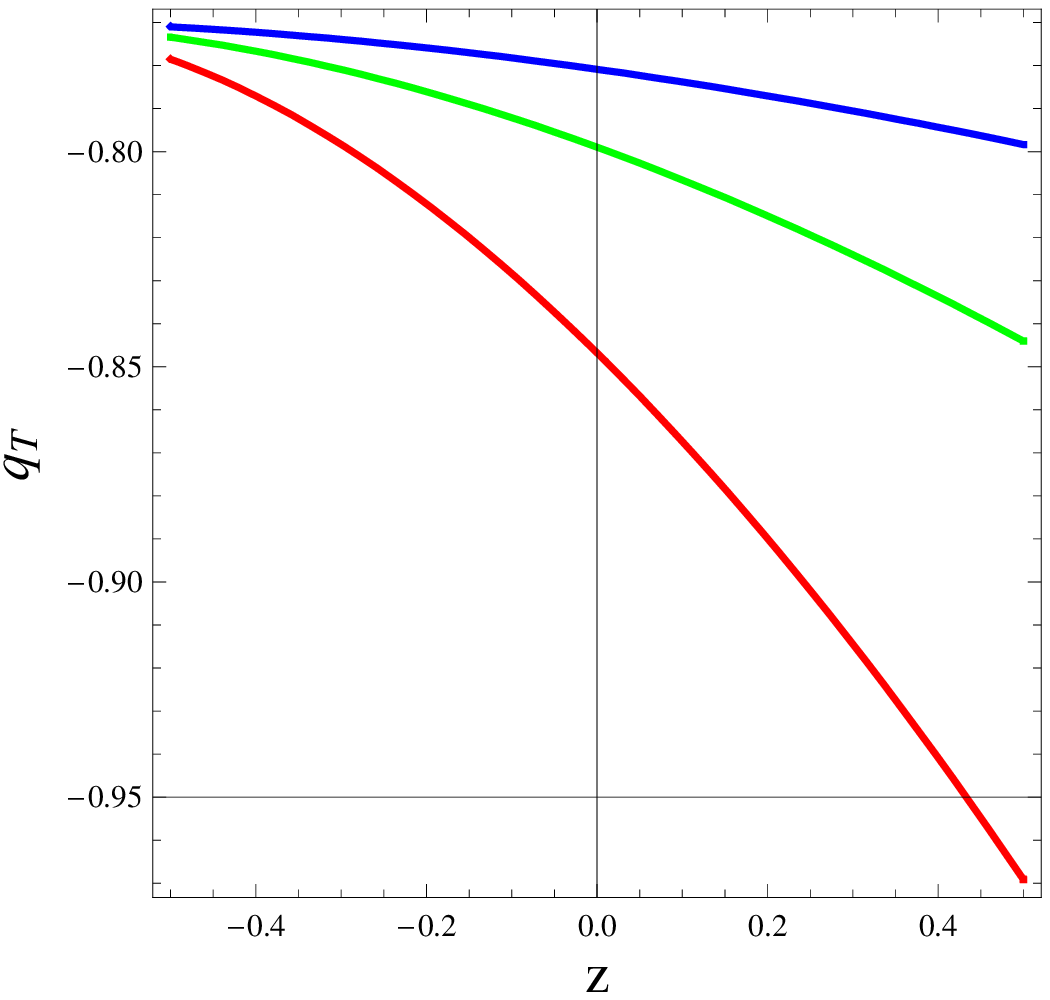}\caption{Deceleration
parameter using Eq. (\ref{expC1hdot}) in Eq. (\ref{qTT}).} \label{13}
\end{minipage}\hspace{0.5cm}\begin{minipage}[b]{0.45\linewidth}
\centering\includegraphics[width=\textwidth]{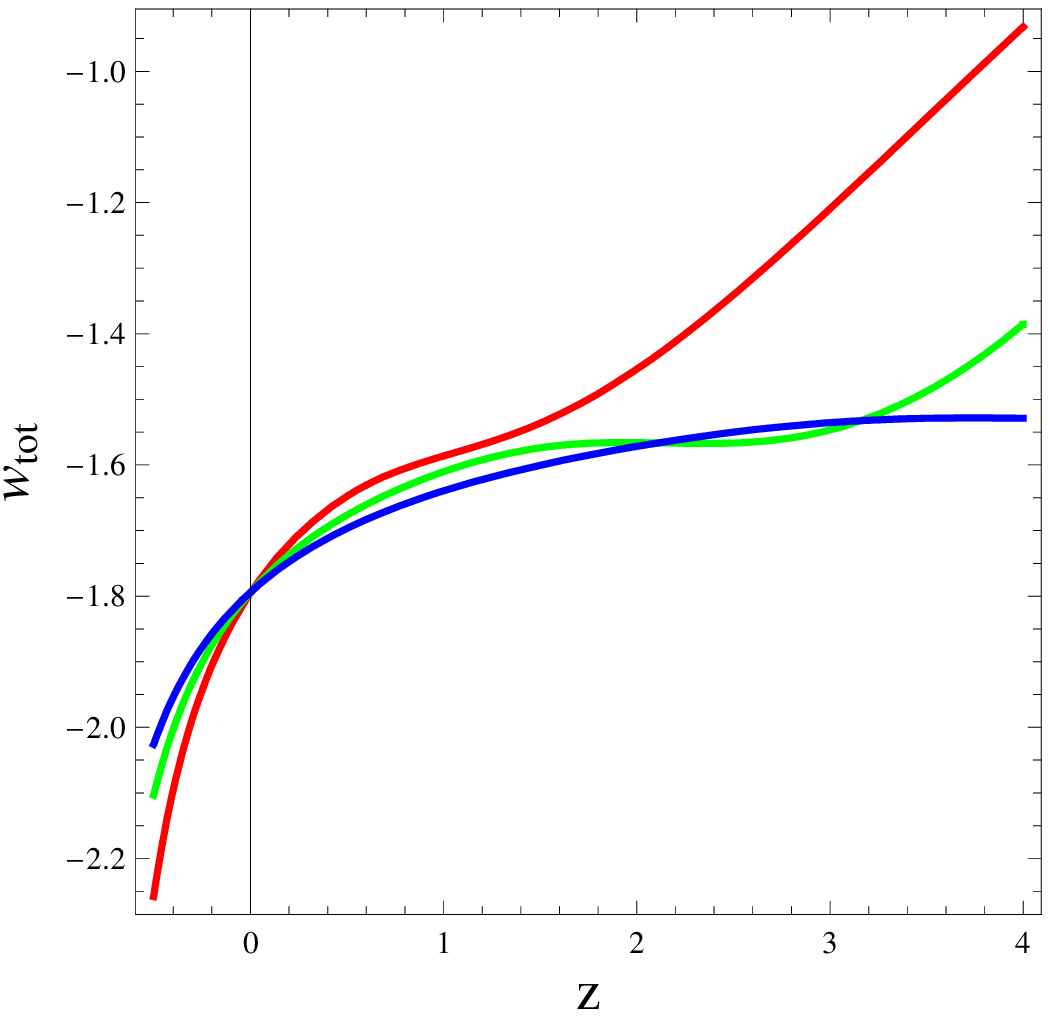}\caption{Plot
of $w_{tot}$ using Eq. (\ref{expC1hdot}) in Eq. (\ref{weffTT}).}
\label{expototalwT}\end{minipage}\end{figure}
\begin{figure}
\centering
\includegraphics[width=18pc]{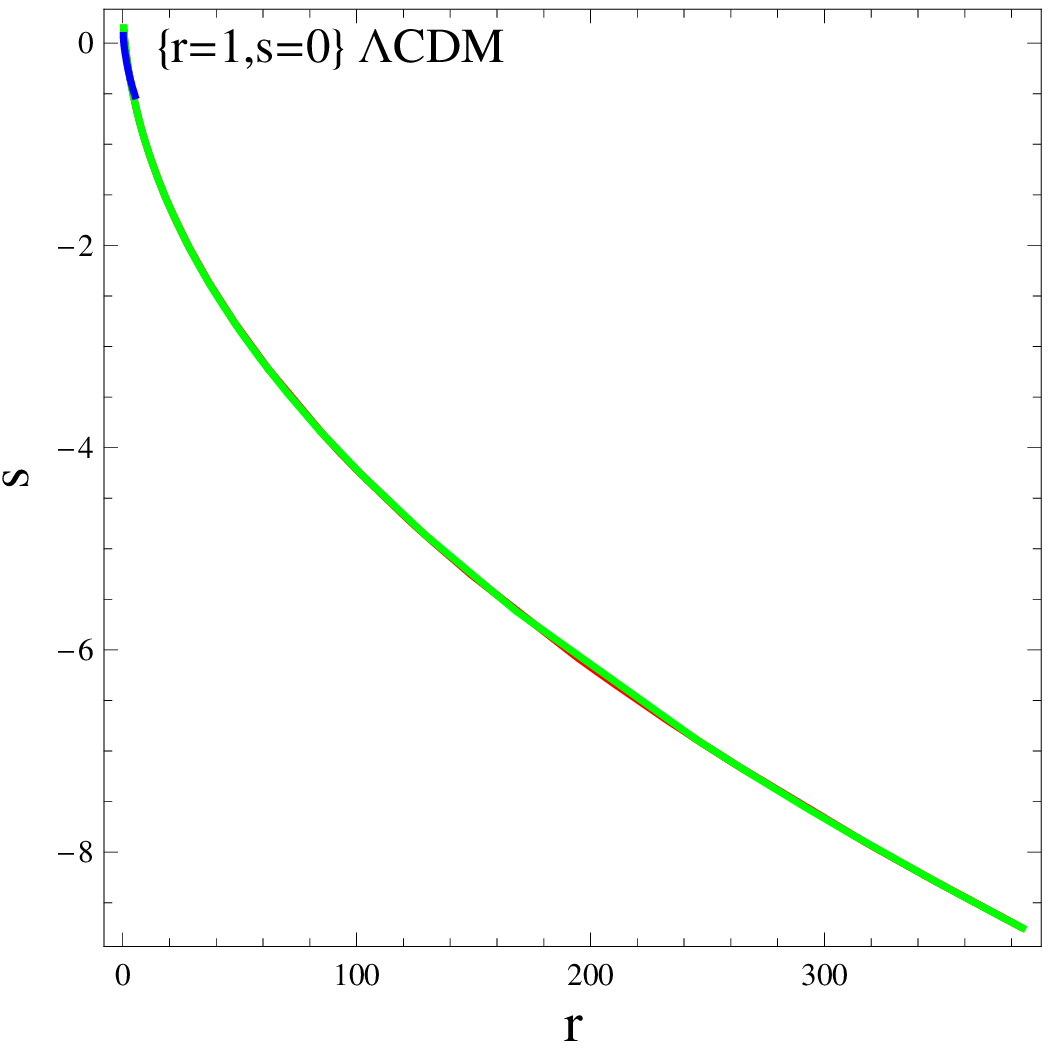}\\
\caption{Statefinder parameters for the choice of $f(T)=\delta \exp (\xi T)$. }\label{state3}
\end{figure}
Fig. \ref{11} shows that $f(T)$ is decreasing with increase in $T$.
It is also observed that after certain stage $f(T)$ is behaving
asymptotically. So, this behavior is in contrary to what happened in
the last two models. Effective torsion parameter $w_T$ displayed in
Fig. \ref{12} behaves like phantom and deceleration parameter
displayed in Fig. \ref{13} makes apparent an ever-accelerating
universe. For $n=6$ (red line), the $w_{total}$ crosses
phantom-divide at $z\approx 3.8$ (Fig. \ref{expototalwT}). However,
for $n=8$ and $10$, $w_{total}$ stays below the phantom-divide. The
statefinder parameters $\{r,s\}$, when plotted in Fig. \ref{state3},
is found to reach $\{r=1,s=0\}_{\Lambda CDM}$, but can not
effectively go beyond it.

\subsection{Comparison with Observational}

By implying different combination of observational schemes at $95\%$
confidence level, Ade et al. \cite{Planck} (Planck data) provided
the following constraints for EoS
\begin{eqnarray*}
w_{DE}&=&-1.13^{+0.24}_{-0.25}~~~~~~~\text{(Planck+WP+BAO)},\\
w_{DE}&=&-1.09\pm0.17,~~~\text{(Planck+WP+Union 2.1)}\\
w_{DE}&=&-1.13^{+0.13}_{-0.14},~~~~~~\text{(Planck+WP+SNLS)},\\
w_{DE}&=&-1.24^{+0.18}_{-0.19},~~~~~~\text{(Planck+WP+$H_0$)}.
\end{eqnarray*}
The trajectories of EoS parameter also favor the following nine-year
WMAP observational data
\begin{eqnarray*}
w_{DE}&=&-1.073^{+0.090}_{-0.089}~~~~~~~\text{(WMAP+eCMB+BAO+$H_0$)},\\
w_{DE}&=&-1.084\pm0.063,~~~\text{(WMAP+eCMB+BAO+$H_0$+SNe)}.
\end{eqnarray*}
It is interesting to mention here that the ranges of EoS parameter
for both cases lie within these observational constraints.

\section{Stability}
\begin{figure}
\centering
\includegraphics[width=18pc]{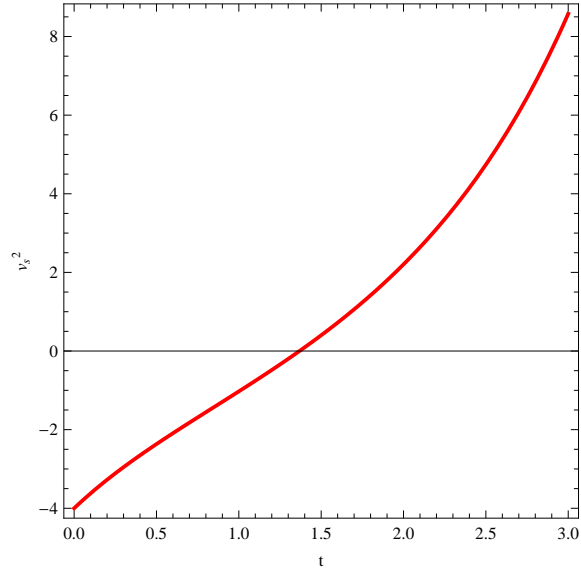}\\
\caption{Plot of Squared speed of sound with specific form of
$H$.}\label{stabH}
\end{figure}
\begin{figure}
\centering
\includegraphics[width=18pc]{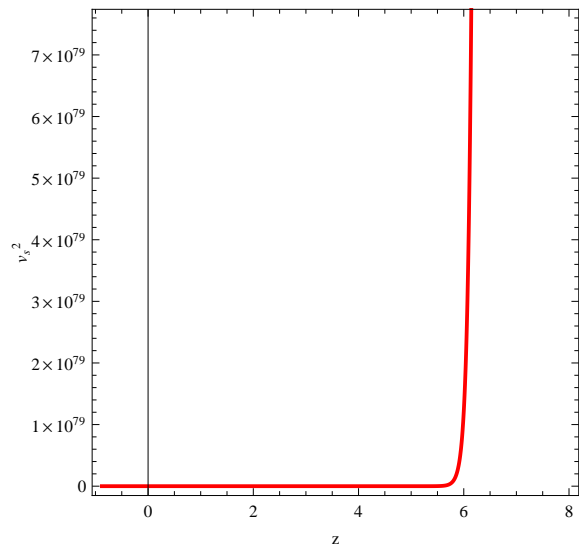}\\
\caption{Plot of Squared speed of sound for power-law form of
$f(T)$.}\label{stabwoH}
\end{figure}
\begin{figure}
\centering
\includegraphics[width=18pc]{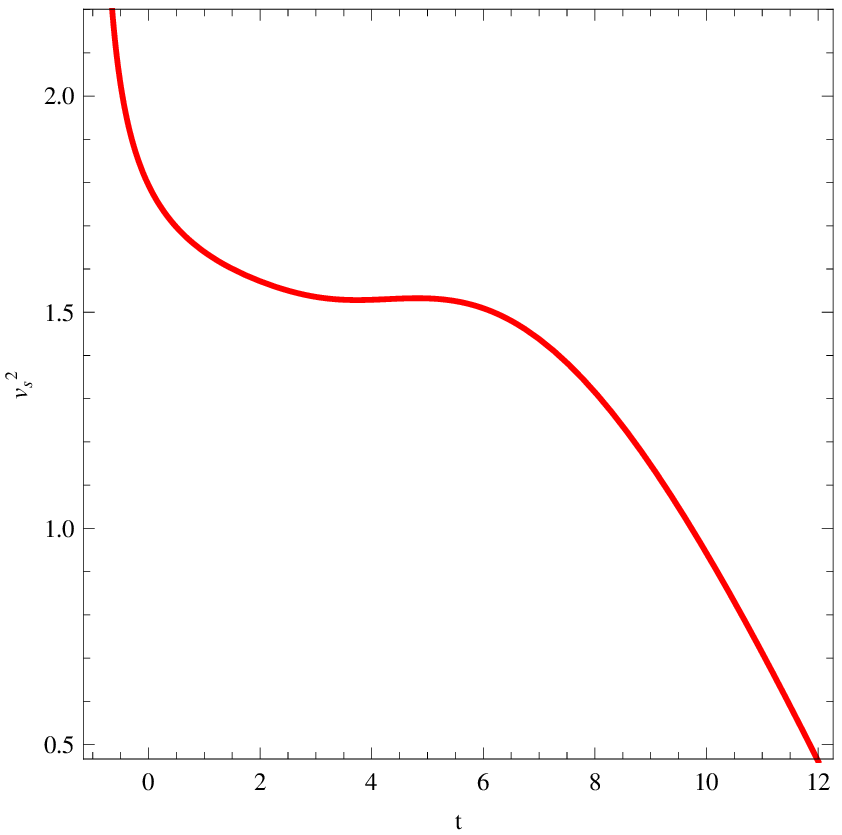}\\
\caption{Plot of Squared speed of sound for exponential form of
$f(T)$.}\label{stabexpo}
\end{figure}
The stability analysis of under consideration models in the present
framework is being discussed in this section. For this purpose, we
consider squared speed of sound which has the following expression
\begin{equation}\label{15-}
v_{s}^2=\frac{\dot{p_T}}{\dot{\rho_T}}.
\end{equation}
The sign of this parameter is very important in order to analyze the
stability of model. This depicts the stable behavior for positive
$v_{s}^2$ while its negativity expresses instability of the under
consideration model. Inserting corresponding expressions and after
some calculations, we can obtain squared speed of sound for all
cases. We draw the graphs versus $t$ for $n=6$ in each case taking
same values for the parameters to discuss the stability of the
reconstructed $f(T)$ model. We provide a discussion about stability
in each case in the following.
\begin{itemize}
\item \textbf{With a specific choice of $H$:}\\Figure \ref{stabH} represents the behavior of $v^2_s$ versus
$t$ for the particular choice of $H$. The graph shows unstable
behavior initially but for a period $t<1.4$. After this interval of
time, squared speed of sound parameter maintain increasing behavior
and becomes positive expressing stability of the model.
\item \textbf{Without any choice of $H$:}\\ In this case, squared speed of sound shows
increasing and positive behavior which exhibits the stability of the
reconstructed model. The corresponding plot is given in Figure
\ref{stabwoH}.
\item \textbf{Exponential Model:}\\ Taking into account the case of exponential
model, we plot the squared speed of sound parameter versus $t$ as
shown in Figure \ref{stabexpo}. The $v^2_s$ represents positively
decreasing behavior establishing stability of the reconstructed
model in this case throughout the time interval.
\end{itemize}

\section{Concluding remarks}

In the present work we have new holographically reconstructed the
polytropic dark energy and this kind of holographic reconstruction
of other dark energy models are already reported in
\cite{recons1,recons2,recons3,recons4}. Viewing $f(T)$ as an
effective description of the underlying theory of DE, and
considering the new holographic polytropic dark energy as pointing
in the direction of the underlying theory of DE, we have studied how
the modified-gravity can describe the new holographic polytropic
dark energy as effective theory of DE. This approach is largely
motivated by \cite{wang,setare}. We have carried out this work
through two approaches. In the first approach we have chosen $H$ as
$H=H_0+\frac{H_1}{t}$ and consequently generated reconstructed
$f(T)$ that is found to tend to $0$ with $T$ tending to $0$ and
thereby satisfying one of the sufficient conditions for a realistic
model \cite{setare}. The effective torsion EoS parameter coming out of this reconstructed
$f(T)$ is found to stay above $-1$ in contradiction to $w_{tot}$
showing a clear transition from quintessence to phantom i.e.
quintom. The deceleration parameter exhibits transition from
decelerated to accelerated phase. The statefinder parameters
$\{r,s\}$ could attain $\Lambda$CDM $\{r=1,s=0\}$ and could go
beyond it. More particularly, it has been apparent from the
statefinder plot that for finite $r$ we have $s\rightarrow -\infty$
that indicates dust phase. Hence this reconstructed $f(T)$ model
interpolates between dust and $\Lambda$CDM phase of the universe.

In the second approach instead of considering any particular form of
the scale factor we have assumed a power-law and exponential
solutions for $f(T)$ as proposed in \cite{BF} and \cite{kaju1}
respectively. Under power-law solution we derived expressions for
$\dot{H}$ in terms of $a$. Thereafter we derived effective torsion
EoS and deceleration parameters and also the statefinder parameters.
For this reconstructed $H$, the $f(T)$ has been found to behave like
the earlier approach that is tending to $0$ as $T$ tends to $0$. As
plotted against redshift $z$, the effective torsion EoS as well as
$w_{tot}$ are found to exhibit phantom-like behavior. The
deceleration parameter is found to stay negative i.e. exhibited
accelerated expansion. Although the statefinder plot could attain
$\Lambda$CDM, no clear attainment of dust phase is apparent. Under
exponential solution of $f(T)$ we derived expressions for $\dot{H}$
in terms of $a$ and subsequently reconstructed $f$ does not tend to
$0$ as $T$ tends to $0$ and hence it does not satisfy the sufficient
condition for realistic model. The effective torsion equation of
state parameter derived this way exhibited phantom-like behavior.
However, $w_{total}$ exhibits a transition from $>-1$ to $<-1$ for
$n=6$.
\begin{figure}
\centering
\includegraphics[width=18pc]{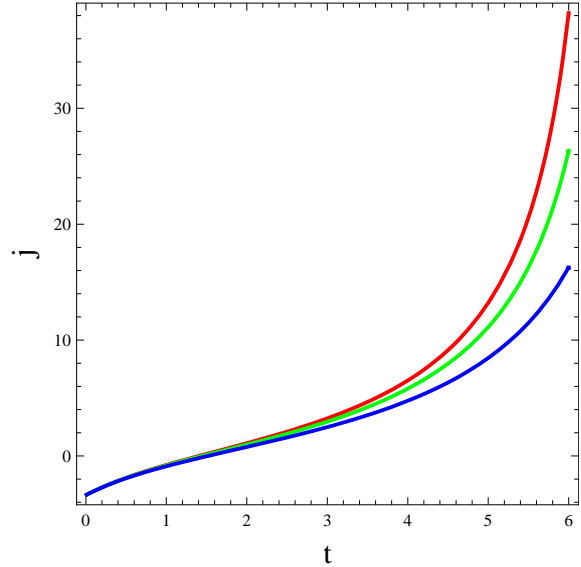}\\
\caption{Jerk parameter plot corresponding to Eq. (\ref{jerkwithout}). }\label{j1plot}
\end{figure}
\begin{figure}
\centering
\includegraphics[width=18pc]{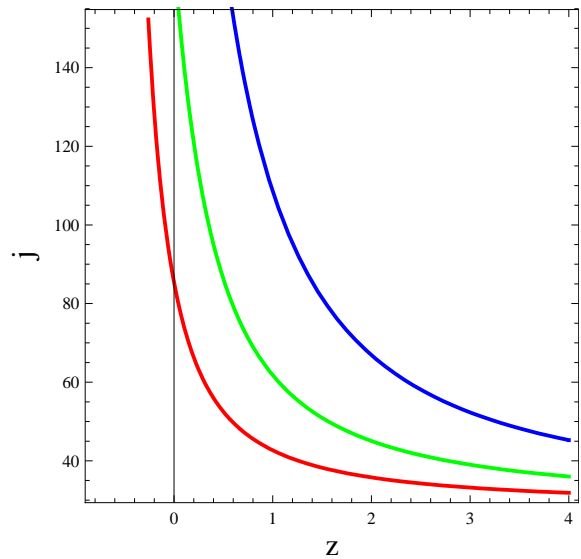}\\
\caption{Jerk parameter plot against redshift $z$ corresponding to Eq. (\ref{jerkwith}). }\label{j2plot}
\end{figure}
We have discussed the stability of the model through squared speed
of sound in all cases. We have obtained a large intervals where
models behave like stable models. Cosmographic parameter $j$, based
on Eqs. (\ref{jerkwithout}) and (\ref{jerkwith}) plotted against $t$
and $z$ in Figs. \ref{j1plot} and \ref{j2plot} respectively show
that for both reconstruction models with and without any choice of
$H$ the jerk parameter $j$ is increasing gradually with evolution of
the universe and remains positive throughout. This observation is
somewhat consistent with the work of \cite{sc}.

In view of the above, although both of the approaches are found to
be somewhat consistent with expected cosmological consequences, the
first approach could be stated to be more acceptable as it could
show a transition from decelerated to accelerated expansion and
could interpolate between dust and $\Lambda$CDM phase of the
universe. Secondly, in the first approach, $w_{total}$ transited
from quintessence to phantom that is found to be consisted with the
outcomes of \cite{kaju1}, where cosmological evolutions of the
equation of state for DE in $f(T)$ gravity was seen to have a
transition of similar nature. However, one major difference between
\cite{kaju1} and the present work lies in the fact that in
\cite{kaju1} the equation of state parameter behaved like quintom
irrespective of exponential, power-law or combined $f(T)$ gravity.
Contrarily, in our present work, the equation of state parameter of
holographic-polytropic gas DE does not necessarily exhibit quintom
behavior.

\section{Acknowledgement}

Constructive suggestions from the reviewers are thankfully
acknowledged by the authors. Visiting associateship of IUCAA, Pune,
India and financial support from DST, Govt of India under project
grant no. SR/FTP/PS-167/2011 are acknowledged by SC.


\begin{thebibliography}{99}

\bibitem{S27} Riess, A.G. et al.: Astron. J. \textbf{116}(1998)1009.

\bibitem{S271} Perlmutter, S. et al.: Astrophys. J.
\textbf{517}(1999)565.

\bibitem{S28} Caldwell, R.R. and Doran, M.: Phys. Rev. D
\textbf{69}(2004)103517.

\bibitem{S281} Koivisto, T. and Mota, D.F.: Phys. Rev. D
\textbf{73}(2006)083502.

\bibitem{S282} Fedeli, C., Moscardini, L. and Bartelmann,
M.: Astron. Astrophys. \textbf{500}(2009)667.

\bibitem{copeland-2006} E. J. Copeland, M. Sami, S. Tsujikawa, Int. J. Mod. Phys. D, \textbf{15}, 1753  (2006).

\bibitem{bambareview} K. Bamba, S. Capozziello, S. Nojiri, S. D. Odintsov, Astrophys.
Space Sci., \textbf{342}, 155 (2012).

\bibitem{DErev} R. R. Caldwell, M. Kamionkowski, Ann. Rev. Nucl. Part. Sci. \textbf{59}, 397 (2009).

\bibitem{nojirireview} S. Nojiri, S. D. Odintsov, Phys. Rep., \textbf{505}, 59 (2011).

\bibitem{star1} V. Sahni, A. A. Starobinsky, Int. J. Mod. Phys. D \textbf{9}, 373 (2000).

\bibitem{satro1} V. Sahni, A. A. Starobinsky, Int. J. Mod. Phys. D \textbf{15}, 2105 (2006).

\bibitem{star2}  H. Motohashi, A. A. Starobinsky, J. Yokoyama, Prog. Theor. Phys.  \textbf{123} , 887 (2010).

\bibitem{star3}  P. J. E. Peebles, B. Ratra, Rev. Mod. Phys. \textbf{75}, 559 (2003).

\bibitem{DE10} S. Nojiri, S. D. Odintsov, O. G. Gorbunova,  J. Phys. A: Math. Gen. \textbf{39}, 6627 (2006).

\bibitem{DE11} A. V. Astashenok, S. Nojiri, S. D. Odintsov, R. J. Scherrer, Phys. Lett. B \textbf{713}, 145 (2012).

\bibitem{DE12} B. Gumjudpai, T. Naskar, M. Sami, S. Tsujikawa, J. Cosmol. Astropart. Phys. \textbf{06}, 007 (2005).

\bibitem{DE13} E. Elizalde, S. Nojiri, S. D. Odintsov, Phys. Rev. D \textbf{70}, 043539 (2004).

\bibitem{DE14} S. Nojiri, S. D. Odintsov, S. Tsujikawa, Phys. Rev. D \textbf{71}, 063004 (2005).

\bibitem{DE15} H. Zhang, Z-H. Zhu, Phys. Rev. D \textbf{73}, 043518 (2006).

\bibitem{DE16} K. Bamba, J. Matsumoto, S. Nojiri, Phys. Rev. D \textbf{85}, 084026 (2012).

\bibitem{DE17} M. Forte, Phys. Rev. D \textbf{90}, 027302 (2014).

\bibitem{NO} Nojiri, S. and Odintsov, S.D.: Int. J. Geom. Meth. Mod. Phys. \textbf{4} 115
(2007).

\bibitem{NO1} Nojiri, S. and Odintsov, S.D.: Phys. Rept. \textbf{505}, 59
(2011).

\bibitem{NO2} Nojiri, S., Odintsov, S. D.: Gen. Relativ. Gravit. \textbf{38}, 1285
(2006).

\bibitem{NO3} Nojiri, S., Odintsov, S. D.: Phys. Lett. B \textbf{631}, 1
(2005).


\bibitem{no} Bamba, K., Nojiri, S. and Odintsov, S.D.: Phys. Lett. B \textbf{725}(2013)368.

\bibitem{no1} Bamba, K. et al.: Astrophys. Space Sci. {\bf 341}(2012)155.

\bibitem{no2} Bamba, K., Jamil, M., Momeni, D. and Myrzakulov, R.: Astrophys.
Space Sci. \textbf{344}(2013)259.


\bibitem{S9} Daouda, M.H., Rodrigues, M.E. and Houndjo, M.J.S.: Eur. Phys. J. C \textbf{72}(2012)1893.

\bibitem{S10} Setare, M.R. and Darabi, F.: Gen. Relativ. Gravit
\textbf{44}(2012)2521.

\bibitem{recons11} S. Nojiri, S. D. Odintsov, Phys. Lett. B \textbf{716}, 377 (2012).

\bibitem{recons21} S. Nojiri, S. D. Odintsov, N. Shirai, J. Cosmol. Astropart. Phys. \textbf{1305}, 020 (2013).

\bibitem{recons31} K. Bamba, S. Nojiri, S. D. Odintsov, Report OCHA-PP-323 (2014).

\bibitem{S11} Farooq, M.U., Jamil, M., Momeni, D. and Myrzakulov,
R.: Can. J. Phys. \textbf{91}(2013)703.

\bibitem{S13} Karami, K. and Abdolmaleki, A.: Res. Astron. Astrophys.
\textbf{13}(2013)757.

\bibitem{SJ} Sharif, M. and Rani, S.: Astrophys. Space Sci.
\textbf{345}(2013)217.

\bibitem{SJ3} Sharif, M. and Rani, S.: Mod. Phys. Lett. A \textbf{29}(2014)1450015.

\bibitem{S7} t'Hooft, G.: gr-qc/9310026.

\bibitem{S71} Susskind, L.: J. Math. Phys. \textbf{36}(1995)6377.

\bibitem{T8z} Bekenstein, J.D.: Phys. Rev. D \textbf{7}(1973)2333.

\bibitem{S8} Cohen, A., Kaplan, D. and Nelson, A.: Phys. Rev. Lett. \textbf{82}(1999)4971.

\bibitem{T11z} Li, M.: Phys. Lett. B \textbf{603}(2004)1.

\bibitem{FS1} Wei, H.: Commun. Theor. Phys.
\textbf{52}(2009)743; Sheykhi, A. and Jamil, M.: Gen. Relativ.
Gravit. \textbf{43}(2011)2661.

\bibitem{FS22} Karami, K. and Abdolmaleki, A.: J. Phys.: Conf. Ser. \textbf{375}
032009 (2012).

\bibitem{FS23} Sharif, M. and Rani, S.: Astrophys. Space Sci.
\textbf{346}(2013)573.

\bibitem{FS24} Sharif, M. and Rani, S.: J. Exp. Theor. Phys.
\textbf{119}(2014)75.

\bibitem{S24} Sahni, V. et al.: J. Exp. Theor. Phys. Lett. \textbf{77}(2003)201.

\bibitem{CP1} A. Aviles, A. Bravetti, S. Capozziello, and O. Luongo
Phys. Rev. D \textbf{87}, 064025 (2013).

\bibitem{CP2} S. Capozziello, O. Luongo, E. N. Saridakis, Phys. Rev. D, \textbf{91}, 124037 (2015).

\bibitem{BF} G. R. Bengochea and R. Ferraro, Phys. Rev. D \textbf{79}, 124019
(2009).

\bibitem{polyplb} K. Karami, S. Ghaffari, Phys. Lett. B  \textbf{688}, 125 (2010).

\bibitem{polyepjc} K. Karami, S. Ghaffari, J. Fehri, Eur. Phys. J. C \textbf{64}, 85 (2009).

\bibitem{granda} L. N. Granda, A. Oliveros, Phys. Lett. B \textbf{669}, 275 (2008).

\bibitem{nojiri} S. Nojiri, S. D. Odintsov, Phys. Rev. D \textbf{74}, 086005 (2006).

\bibitem{recons1} K. Karami, J. Fehri, Phys. Lett. B \textbf{684}, 61 (2010).

\bibitem{recons2} U. Debnath, Eur. Phys. J. Plus \textbf{129}, 272 (2014).

\bibitem{recons3} A. Sheykhi, Phys. Rev. D \textbf{84}, 107302 (2011).

\bibitem{recons4} S. Chattopadhyay, A. Pasqua, M. Khurshudyan, Eur. Phys. J. C \textbf{74}, 3080 (2014).

\bibitem{wang} W. Yang, Y. Wu, L. Song, Y. Su, J. Li, D. Zhang, X. Wang, Mod.
Phys. Lett. A \textbf{26}, 191 (2011).

\bibitem{setare} M. R. Setare, Int. J. Mod. Phys. D, \textbf{17}, 2219 (2008).

\bibitem{kaju1} K. Bamba, C-Q. Geng, C-C. Lee, L-W. Luo, JCAP \textbf{021} 1101
(2011).

\bibitem{sc} S. Pan, S. Chakraborty, Int. J. Mod. Phys. D \textbf{23}, 1450092 (2014).

\bibitem{Planck} Ade, P.A.R., et al.: Astronom. Astrophys.
\textbf{571}(2014)A16.
\end{thebibliography}
\end{document}